\documentclass{aa}
\usepackage{color}
\usepackage{graphicx}
\usepackage{lscape}
\usepackage{natbib}
\usepackage[scaled]{helvet}
\usepackage[varg]{txfonts}
\usepackage{url}
\usepackage{xspace}
\bibpunct{(}{)}{;}{a}{}{,}
\newcounter{Rco}
\newcommand{\ionw}[3]{\mbox{\ion{#1}{#2}~$\lambda\,#3\,\mathrm{\AA}$}\xspace}
\newcommand{\ionww}[3]{\mbox{\ion{#1}{#2}~$\lambda\lambda\,#3\,\mathrm{\AA}$}\xspace}
\newcommand{\Ionst}[1]{\setcounter{Rco}{#1}\Roman{Rco}}
\newcommand{\Ion}[2]{\mbox{#1\,{\scriptsize\Ionst{#2}}}}
\newcommand{\Ionw}[3]{\mbox{#1\,{\scriptsize\Ionst{#2}}~$\lambda\,#3$\,\AA}\xspace}
\newcommand{\Ionwf}[3]{\mbox{[#1\,{\scriptsize\Ionst{#2}}]~$\lambda\,#3$\,\AA}\xspace}
\newcommand{\Ionww}[3]{\mbox{#1\,{\scriptsize\Ionst{#2}}~$\lambda\lambda\,#3$\,\AA}\xspace}
\newcommand{\Jonw}[3]{\mbox{\ion{#1}{#2}~$\lambda\,#3$\,\AA}\xspace}
\newcommand{\Jonww}[3]{\mbox{\ion{#1}{#2}~$\lambda\lambda\,#3$\,\AA}\xspace}
\newcommand{\ea}{et al\@. }
\newcommand{\logg}{\mbox{$\log g$}\xspace}
\newcommand{\loggw}[1]{\mbox{$\log g\hspace{-0.5mm} =\hspace{-0.5mm}  #1$}}
\newcommand{\kK}{\mathrm{kK}}
\newcommand{\ab}[1]{\mbox{Fig.\,\ref{#1}}}
\newcommand{\sA}[1]{\mbox{(Fig.\,\ref{#1})}}
\newcommand{\ratio}[2]{\mbox{$n_{\rm #1}/n_{\rm #2}$}}
\newcommand{\ratiow}[3]{\mbox{$n_{\rm #1}/n_{\rm #2}\hspace{-0.5mm} = \hspace{-0.5mm} #3$}}
\newcommand{\se}[1]{\mbox{Sect.\,\ref{#1}}}
\newcommand{\sga}{\raisebox{-0.10em}{$\stackrel{>}{{\mbox{\tiny $\sim$}}}$}}
\newcommand{\sK}[1]{\mbox{(Sect.\,\ref{#1})}}
\newcommand{\sla}{\raisebox{-0.10em}{$\stackrel{<}{{\mbox{\tiny $\sim$}}}$}}
\newcommand{\spm}{\mbox{\raisebox{0.20em}{{\tiny \hspace{0.2mm}\mbox{$\pm$}\hspace{0.2mm}}}}}
\newcommand{\ta}[1]{\mbox{Tab.\,\ref{#1}}}
\newcommand{\sT}[1]{\mbox{(Tab.\,\ref{#1})}}
\newcommand{\Teff}{\mbox{$T_\mathrm{eff}$}\xspace}
\newcommand{\Teffw}[1]{\mbox{$\Teff\hspace{-0.5mm} =\hspace{-0.5mm} #1 \,\mathrm{K}$}}
\newcommand{\ebv}{\mbox{$E_\mathrm{B-V}$}}
\newcommand{\ebvw}[1]{\mbox{$\ebv\hspace{-0.5mm} =\hspace{-0.5mm} #1$}}
\newcommand{\deh}{\mbox{$N_\ion{D}{i}$}}
\newcommand{\nh}{\mbox{$N_\ion{H}{i}$}}
\newcommand{\nhw}[1]{\mbox{$\nh\hspace{-0.5mm} =\hspace{-0.5mm} #1\, \mathrm{cm}^{-2}$}}
\newcommand{\vrad}{\mbox{$v_\mathrm{rad}$}}
\newcommand{\vradw}[1]{\mbox{$\vrad = \hspace{-0.5mm} #1\, \mathrm{km\,sec}^{-1}$}}
\newcommand{\Msol}{$M_\odot$\xspace}
\newcommand{\lmspr}{\hbox{}\hspace{+1.2cm}}
\newcommand{\mmspr}{\hbox{}\hspace{+0.7cm}}
\newcommand{\smspr}{\hbox{}\hspace{+2.5mm}}
\newcommand{\smspl}{\hbox{}\hspace{+0.1mm}}
\newcommand{\gb}{\object{G191$-$B2B}\xspace}
\newcommand{\re}{\object{RE\,0503$-$289}\xspace}
\begin{document}

\title{Stellar laboratories }
\subtitle{IX. New \ion{Se}{v},
                  \ion{Sr}{iv -- vii},
                  \ion{Te}{vi}, and
                  \ion{I}{vi}
              oscillator strengths and the Se, Sr, Te, and I abundances in the hot white dwarfs \gb and \re
           \thanks
           {Based on observations with the NASA/ESA Hubble Space Telescope, obtained at the Space Telescope Science 
            Institute, which is operated by the Association of Universities for Research in Astronomy, Inc., under 
            NASA contract NAS5-26666.
           }\fnmsep
           \thanks
           {Based on observations made with the NASA-CNES-CSA Far Ultraviolet Spectroscopic Explorer.
           }\fnmsep
           \thanks
           {Tables \ref{tab:sev:loggf} to \ref{tab:ivi:loggf} are only available via the
            German Astrophysical Virtual Observatory (GAVO) service TOSS (http://dc.g-vo.org/TOSS).
           }
         }

\titlerunning{Stellar laboratories: New \ion{Sr}{iv -- vii} oscillator strengths}

\author{T\@. Rauch\inst{1}
        \and
        P\@. Quinet\inst{2,3}
        \and
        M\@. Kn\"orzer\inst{1}
        \and
        D\@. Hoyer\inst{1}
        \and
        K\@. Werner\inst{1}
        \and
        J\@. W\@. Kruk\inst{4}
        \and
        M\@. Demleitner\inst{5}
        }

\institute{Institute for Astronomy and Astrophysics,
           Kepler Center for Astro and Particle Physics,
           Eberhard Karls University,
           Sand 1,
           72076 T\"ubingen,
           Germany \\
           \email{rauch@astro.uni-tuebingen.de}
           \and
           Physique Atomique et Astrophysique, Universit\'e de Mons -- UMONS, 7000 Mons, Belgium
           \and
           IPNAS, Universit\'e de Li\`ege, Sart Tilman, 4000 Li\`ege, Belgium
           \and
           NASA Goddard Space Flight Center, Greenbelt, MD\,20771, USA
           \and
           Astronomisches Rechen-Institut (ARI), Centre for Astronomy of Heidelberg University, M\"onchhofstra\ss e 12-14, 69120 Heidelberg, Germany}

\date{Received 2 January 2017; accepted 19 June 2017}

\abstract {To analyze spectra of hot stars, advanced non-local thermodynamic equilibrium (NLTE) 
           model-atmosphere techniques are mandatory. Reliable atomic data is crucial for the calculation of such model atmospheres.
          }
          {We aim to calculate new \ion{Sr}{iv-vii} oscillator strengths to identify for the first time 
           Sr spectral lines in hot white dwarf (WD) stars and to determine the photospheric Sr abundances.
           To measure the abundances of Se, Te, and I in hot WDs, we aim to compute new \ion{Se}{v}, \ion{Te}{vi}, and \ion{I}{vi} 
           oscillator strengths.
          }
          {To consider radiative and collisional bound-bound transitions of
           \ion{Se}{v},
           \ion{Sr}{iv - vii},
           \ion{Te}{vi}, and 
           \ion{I}{vi} in our NLTE atmosphere models,
           we calculated oscillator strengths for these ions.
          }
          {We newly identified
            four \ion{Se}{v}, 
           23 \ion{Sr}{v},
            1 \ion{Te}{vi}, and
            three \ion{I}{vi} 
           lines in the ultraviolet (UV) spectrum of \re.
           We measured a photospheric 
           Sr abundance of $6.5^{+3.8}_{-2.4} \times 10^{-4}$ (mass fraction, 9\,500 - 23\,800 times solar).
           We determined the abundances of
           Se ($1.6^{+0.9}_{-0.6} \times 10^{-3}$, 8\,000 - 20\,000),
           Te ($2.5^{+1.5}_{-0.9} \times 10^{-4}$, 11\,000 - 28\,000), and
           I  ($1.4^{+0.8}_{-0.5} \times 10^{-5}$,  2\,700 -  6\,700).
           No Se, Sr, Te, and I line was found in the UV spectra of \gb and we could determine only upper abundance limits
           of approximately 100 times solar.
          }
          {All identified \ion{Se}{v}, \ion{Sr}{v}, \ion{Te}{vi}, and \ion{I}{vi} lines in the UV spectrum of \re 
           were simultaneously well reproduced with our newly calculated oscillator strengths.
          }


\maketitle

\section{Introduction}
\label{sect:intro}

\begin{figure}
   \resizebox{\hsize}{!}{\includegraphics{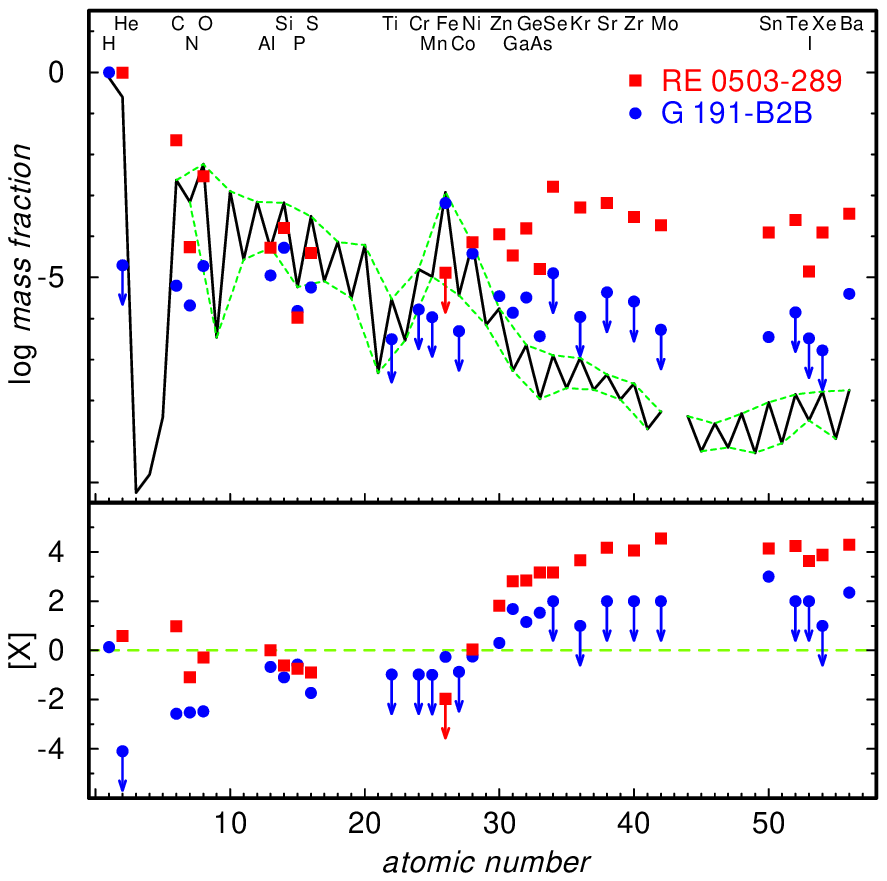}}
    \caption{Solar abundances \citep[thick line; the dashed lines
             connect the elements with even and with odd atomic number]{asplundetal2009,scottetal2015a,scottetal2015b,grevesseetal2015}
             compared with the determined photospheric abundances of 
             \re \citep[red squares,][and this work]{dreizlerwerner1996, rauchetal2012ge, rauchetal2014zn, rauchetal2014ba, rauchetal2015ga, rauchetal2016mo, rauchetal2016kr, rauchetal2016zr}.
             The uncertainties of the WD abundances are about 0.2\,dex in general. Arrows indicate upper limits.
             Top panel: Abundances given as logarithmic mass fractions.
             Bottom panel: Abundance ratios to respective solar values, 
                           [X] denotes log (fraction\,/\,solar fraction) of species X.
                           The dashed, green line indicates solar abundances.
            }
   \label{fig:X}
\end{figure}

Recent spectral analyses \citep[cf.,][]{rauchetal2016zr} of high-resolution UV spectra of the
helium-rich (DO-type) white dwarf (WD) \re
\citep[\object{RX\,J0503.9$-$2854}, \object{WD\,0501+527},][]{mccooksion1999,mccooksion1999cat}
revealed strongly enriched trans-iron elements (atomic numbers $Z \ge 30$) in its photosphere
(Fig.\,\ref{fig:X}). Efficient radiative levitation 
\citep{rauchetal2016mo} in this hot WD 
\citep[effective temperature \Teffw{70\,000 \pm 2000}, 
       surface gravity $\log\,(g\,/\,\mathrm{cm\,s^{-2}}) = 7.5 \pm 0.1$,][]{rauchetal2016kr}
can increase abundances by more than 4\,dex compared with solar values.
In the cooler \citep[\Teffw{60\,000 \pm 2000}, \loggw{7.6 \pm 0.05},][]{rauchetal2013}, hydrogen-rich (DA-type) WD
\gb \citep[\object{WD\,0501+527},][]{mccooksion1999,mccooksion1999cat},
the radiative levitation is able to retain only a factor of $\approx 100$ fewer trans-iron elements
in the photosphere than in \re (Fig.\,\ref{fig:X}).

The search for signatures of trans-iron elements in the spectra of \re and \gb was initiated by the 
discovery of Ga, Ge, As, Se, Kr, Mo, Sn, Te, I, and Xe lines in \re \citep{werneretal2012}.
Subsequent calculations of transition probabilities allowed reliable abundance determinations of
Zn (atomic number $Z = 30$), 
Ga (31), 
Ge (32), 
Kr (36), 
Zr (40),
Mo (42),
Xe (54), and 
Ba (56) \citep[e.g.,][and references therein]{rauchetal2016zr}.
Based on the wavelengths provided by the Atomic Spectra Database 
(ASD\footnote{\url{http://physics.nist.gov/PhysRefData/ASD/lines_form.html}}) of the
National Institute of Standards and Technology (NIST),
we have identified some strong lines of strontium (38), an element that was hitherto not detected in hot WDs.
For an identification of other, weaker Sr lines and a subsequent abundance analysis, we decided to calculate
new \ion{Sr}{iv-vii} transition probabilities. 

The paper is organized as follows.
We briefly introduce the UV spectra in Sect.\,\ref{sect:observation}. 
Our model atmospheres, the atomic data as well as the transition-probability calculations are described in Sect.\,\ref{sect:models}.
Here, we have included the calculation of new transition probabilities for \ion{Se}{v}, \ion{Te}{vi}, and \ion{I}{vi}
because these are the last three elements (34, 52, and 53, respectively), 
that were previously identified by \citet{werneretal2012} in the spectrum of \re.
The line identification and abundance analysis then follows in Sect.\,\ref{sect:results}.

\section{Observations}
\label{sect:observation}

Our analysis is based on UV spectroscopy that was performed with 
the Far Ultraviolet Spectroscopic Explorer 
(FUSE, $910\,\mathrm{\AA} < \lambda < 1190\,\mathrm{\AA}$, resolving power $R \approx 20\,000$) and
the Hubble Space Telescope / Space Telescope Imaging Spectrograph 
(HST/STIS, $1144\,\mathrm{\AA} < \lambda < 1709\,\mathrm{\AA}$, $R \approx 45\,800$).
The spectra are described in detail in \citet{hoyeretal2017}.
The observed spectra shown here were shifted to rest wavelengths, using
$v_\mathrm{rad} = 24.56\,\mathrm{km\,s^{-1}}$ for \gb \citep{lemoineetal2002} and 
                 $25.8\,\mathrm{km\,s^{-1}}$ for \re \citep{hoyeretal2017}.
To compare them with our synthetic spectra, the latter were convolved with Gaussians to 
model the respective instruments' resolving power.

\section{Model atmospheres and atomic data}
\label{sect:models}

To calculate model atmospheres for our analysis, we used the T\"ubingen Model-Atmosphere Package
\citep[TMAP\footnote{\url{http://astro.uni-tuebingen.de/~TMAP}},][]{werneretal2003,tmap2012}.
These models are plane-parallel, chemically homogeneous, and in hydrostatic and radiative 
equilibrium. TMAP considers non-local thermodynamic equilibrium (NLTE). 
More details are given by \citet{rauchetal2016kr}.
We include opacities 
of 
H$^\mathrm{G}$,
He,           
C,            
N,            
O,            
Al,           
Si,           
P,            
S,            
Ca,           
Sc,           
Ti,           
V,            
Cr,           
Mn,           
Fe,           
Co,           
Ni,           
Zn,           
Ga,           
Ge,           
As,           
Se,           
Kr$^\mathrm{R}$,
Sr,           
Zr,           
Mo,           
Sn,           
Te,           
I,           
Xe$^\mathrm{R}$, and
Ba ($^\mathrm{G}$: only in \gb models,
    $^\mathrm{R}$: only in \re models).
Model atoms for all species with $Z < 20$ are compiled from the T\"ubingen Model Atom Database (TMAD).
For the iron-group elements (Ca - Ni, $20 \leq Z \leq 28$), model atoms were constructed with a statistical approach
by calculating so-called super levels and super lines \citep[IrOnIc code,][]{rauchdeetjen2003} with the T\"ubingen Iron-Group Opacity -- 
IrOnIc WWW Interface \citep{muellerringatPhD2013}. For trans-iron elements ($Z \geq 29$), we transferred their atomic data 
into Kurucz-formatted files \citep[cf.,][]{rauchetal2015ga}, and followed the same statistical method.
The Se, Sr, Te, and I model-atom statistics are given in Table\,\ref{tab:ironic}.

\begin{table}\centering
\caption{Statistics of the  
         \ion{Se}{v},
         \ion{Sr}{iv - vii},
         \ion{Te}{vi}, and 
         \ion{I}{vi} atomic levels and line transitions from
         Tables\,\ref{tab:sev:loggf} - \ref{tab:ivi:loggf}.
        } 
\label{tab:ironic}
\begin{tabular}{r@{\,}lcccc}
\hline
\hline
\multicolumn{2}{c}{Ion}        & Atomic levels & Lines & Super levels & Super lines \\
\hline
Se&{\sc v}                     &            46 &   310 &            7 &          19 \\
Sr&{\sc iv}                    &           254 &  7578 &            7 &          21 \\
Sr&{\sc v}                     &           130 &  2022 &            7 &          19 \\
Sr&{\sc vi}                    &            22 &    70 &            7 &          10 \\
Sr&{\sc vii}                   &            19 &    46 &            7 &          10 \\
Te&{\sc vi}                    &            30 &   178 &            7 &          12 \\
 I&{\sc vi}                    &            38 &   197 &            7 &          15 \\
\hline
\end{tabular}
\end{table}

\paragraph{}
New sets of oscillator strengths and transition 
probabilities for the \ion{Se}{v}, \ion{Sr}{iv-vii}, \ion{Te}{vi}, and \ion{I}{vi} ions were computed 
using the pseudo-relativistic Hartree-Fock (HFR) approach of \citet{cowan1981} 
modified for including core-polarization effects, giving rise to the HFR+ CPOL 
method, as described by \citet{quinetetal1999,quinetetal2002}.
For each ion, this method was
combined with a semi-empirical least-squares fit of radial energy
parameters to minimize the differences between computed and
available experimental energy levels.

\paragraph{\ion{Se}{v}:\hspace{10mm}}
The 
4s$^2$, 
4p$^2$, 
4d$^2$, 
4f$^2$, 
4s4d, 
4s5d, 
4s6d, 
4s5s,
4s6s, 
4p4f, 
4p5f, 
4p6f, 
4d5s, 
4d6s, 
4d5d, and
4d6d even-parity
configurations and the 
4s4p, 
4s5p, 
4s6p, 
4s4f, 
4s5f, 
4s6f, 
4p4d, 
4p5d,
4p6d, 
4p5s, 
4p6s, 
4d4f, and
4d5f odd-parity configurations were explicitly
included in the physical model. Core-polarization effects were
estimated by assuming a Ni-like \ion{Se}{vii} ionic core with a
core-polarizability $\alpha$$_d$ of 0.36\,a.u., as reported by 
\citet{johnsonetal1983}, and a cut-off radius, $r_\mathrm{c}$ equal to 0.62\,a.u.,
corresponding to the HFR mean radius of the outermost core orbital
(3d). Using the experimental energy levels published by \citet{churilovjoshi1995}, 
the radial integrals characterizing the 
4s$^2$, 
4p$^2$,
4s4d, 
4s5d, 
4s5s, 
4s6s, 
4s4p, 
4s5p, 
4s4f, 
4p4d, and 
4p5s configurations
were fitted. This semi-empirical adjustment allowed us to reduce the
average deviations between calculated and measured energies to 8\,cm$^{-1}$ and 
219\,cm$^{-1}$ for even and odd parities, respectively.

\paragraph{\ion{Sr}{iv}:\hspace{10mm}}
We considered interaction among the configurations 
4s$^2$4p$^5$, 
4s$^2$4p$^4$5p, 
4s$^2$4p$^4$6p,
4s$^2$4p$^4$4f, 
4s$^2$4p$^4$5f, 
4s$^2$4p$^4$6f, 
4s$^2$4p$^4$6h,
4s4p$^5$4d, 
4s4p$^5$5d, 
4s4p$^5$6d, 
4s4p$^5$5s, 
4s4p$^5$6s,
4s4p$^5$5g, 
4s4p$^5$6g, 
4s$^2$4p$^3$4d$^2$, 
4s$^2$4p$^3$4f$^2$, and
4p$^6$4f for the odd parity, and 
4s4p$^6$, 
4s$^2$4p$^4$4d,
4s$^2$4p$^4$5d, 
4s$^2$4p$^4$6d, 
4s$^2$4p$^4$5s, 
4s$^2$4p$^4$6s,
4s$^2$4p$^4$7s, 
4s$^2$4p$^4$5g, 
4s$^2$4p$^4$6g, 
4s4p$^5$4f,
4s4p$^5$5f, 
4s4p$^5$6f, 
4s4p$^5$5p, 
4s4p$^5$6p, 
4s4p$^5$6h, 
4p$^6$4d, and
4p$^6$5s for the even parity. The core-polarization parameters were
the dipole polarizability of a Ni-like \ion{Sr}{ix} ionic core as reported by
\citet{johnsonetal1983}, that is, $\alpha$$_d$ = 0.13 a.u., and the cut-off
radius corresponding to the HFR mean value $<$$r$$>$ of the outermost
core orbital (3d), i.e., $r_\mathrm{c} = 0.49\,\mathrm{a.u.}$  Using experimental energy
levels compiled by \citet{sansonetti2012}, the radial integrals (average
energy, Slater, spin-orbit and effective interaction parameters) of
4p$^5$, 
4p$^4$5p, 
4p$^4$6p, 
4p$^4$4f, 
4p$^4$5f, 
4p$^4$6h, 
4s4p$^5$4d,
4s4p$^6$, 
4p$^4$4d, 
4p$^4$5d, 
4p$^4$6d,  
4p$^4$5s, 
4p$^4$6s, 
4p$^4$7s,
4p$^4$5g, and 
4p$^4$6g configurations were optimized by a least-squares
fitting procedure in which the mean deviations with experimental data
were found to be equal to 145\,cm$^{-1}$ for the odd parity and 150\,cm$^{-1}$ for the even parity.

\paragraph{\ion{Sr}{v}:\hspace{10mm}}
The HFR method was used with, as interacting configurations,
4s$^2$4p$^4$, 
4s$^2$4p$^3$5p, 
4s$^2$4p$^3$6p, 
4s$^2$4p$^3$4f,
4s$^2$4p$^3$5f, 
4s$^2$4p$^3$6f, 
4s$^2$4p$^3$6h, 
4s4p$^4$4d,
4s4p$^4$5d, 
4s4p$^4$6d, 
4s4p$^4$5s, 
4s4p$^4$6s, 
4s4p$^4$5g,
4s4p$^4$6g, 
4s$^2$4p$^2$4d$^2$, 
4s$^2$4p$^2$4f$^2$, 
4p$^6$, and
4p$^5$4f
for the even parity, and 
4s4p$^5$, 
4s$^2$4p$^3$4d, 
4s$^2$4p$^3$5d,
4s$^2$4p$^3$6d, 
4s$^2$4p$^3$5s, 
4s$^2$4p$^3$6s, 
4s$^2$4p$^3$5g,
4s$^2$4p$^3$6g, 
4s4p$^4$4f, 
4s4p$^4$5f, 
4s4p$^4$6f, 
4s4p$^4$5p,
4s4p$^4$6p, 
4s4p$^4$6h, 
4p$^5$4d, and
4p$^5$5s for the odd
parity. Core-polarization effects were estimated using the same
$\alpha_\mathrm{d}$ and $r_\mathrm{c}$ values as those considered in \ion{Sr}{iv}. The radial
integrals corresponding to 
4p$^4$, 
4p$^3$5p, 
4s4p$^5$, 
4p$^3$4d,
4p$^3$5d, 
4p$^3$5s, and 
4p$^3$6s were adjusted to reproduce at best the
experimental energy levels tabulated by \citet{sansonetti2012}. We note that
the few levels reported by this author as belonging to the 4p$^3$4f
and 4p$^3$5f configurations were not included in the fitting process
because it was found that most of those levels were strongly mixed
with states of experimentally unknown configurations, such as
4s4p$^4$4d, 4p$^3$6p, 4p$^2$4d$^2$, and 4s4p$^4$5s. It was then
extremely difficult to establish an unambiguous correspondence between
the calculated and experimental energies. For the levels considered in
our semi-empirical adjustment, we found mean deviations equal to 
138\,cm$^{-1}$ and 231\,cm$^{-1}$ in even and odd parities, respectively.

\paragraph{\ion{Sr}{vi}:\hspace{10mm}}
The configurations included in the HFR model
were 
4s$^2$4p$^3$, 
4s$^2$4p$^2$5p, 
4s$^2$4p$^2$6p, 
4s$^2$4p$^2$4f,
4s$^2$4p$^2$5f, 
4s$^2$4p$^2$6f, 
4s$^2$4p$^2$6h, 
4s4p$^3$4d,
4s4p$^3$5d, 
4s4p$^3$6d, 
4s4p$^3$5s, 
4s4p$^3$6s, 
4s4p$^3$5g,
4s4p$^3$6g, 
4s$^2$4p4d$^2$, 
4s$^2$4p4f$^2$, 
4p$^5$, and
4p$^4$4f for the odd parity, and 
4s4p$^4$, 
4s$^2$4p$^2$4d, 
4s$^2$4p$^2$5d,
4s$^2$4p$^2$6d, 
4s$^2$4p$^2$5s, 
4s$^2$4p$^2$6s, 
4s$^2$4p$^2$5g,
4s$^2$4p$^2$6g, 
4s4p$^3$4f, 
4s4p$^3$5f, 
4s4p$^3$6f, 
4s4p$^3$5p,
4s4p$^3$6p, 
4s4p$^3$6h, 
4p$^4$4d, and
4p$^4$5s for the even parity. 
The same core-polarization parameters as those used for \ion{Sr}{iv} were
considered while the fitting process was performed with the few
experimental energy levels listed in the compilation of \citet{sansonetti2012}
for optimizing the radial parameters of 4p$^3$, 4s4p$^4$, and
4p$^2$5s configurations, leading to mean deviations equal to 13\,cm$^{-1}$ (odd parity) and 32\,cm$^{-1}$ (even parity).

\paragraph{\ion{Sr}{vii}:\hspace{10mm}}
A model similar to that of \ion{Sr}{vi} was used, for which the
4s$^2$4p$^2$, 
4s$^2$4p5p, 
4s$^2$4p6p, 
4s$^2$4p4f, 
4s$^2$4p5f,
4s$^2$4p6f, 
4s$^2$4p6h, 
4s4p$^2$4d, 
4s4p$^2$5d, 
4s4p$^2$6d,
4s4p$^2$5s, 
4s4p$^2$6s, 
4s4p$^2$5g, 
4s4p$^2$6g, 
4s$^2$4d$^2$,
4s$^2$4f$^2$, 
4p$^4$, and 
4p$^3$4f even-parity configurations and the
4s4p$^3$, 
4s$^2$4p4d, 
4s$^2$4p5d, 
4s$^2$4p6d, 
4s$^2$4p5s, 
4s$^2$4p6s,
4s$^2$4p5g, 
4s$^2$4p6g, 
4s4p$^2$4f, 
4s4p$^2$5f, 
4s4p$^2$6f,
4s4p$^2$5p, 
4s4p$^2$6p, 
4s4p$^2$6h, 
4p$^3$4d, and 
4p$^3$5s odd-parity
configurations were explicitly included in the HFR model. Here also,
we used the same core-polarization parameters as those considered for
\ion{Sr}{iv}. The semi-empirical optimization process was carried out to
adjust the radial parameters in 4p$^2$, 4s4p$^3$, and 4p5s with the
experimental energy levels taken from \citet{sansonetti2012} giving rise to
average deviations of 0\,cm$^{-1}$ and 247\,cm$^{-1}$ for even and odd
parities, respectively.

\paragraph{\ion{Te}{vi}:\hspace{10mm}}
The configuration interaction was considered among
the following configurations: 
4d$^{10}$5s, 
4d$^{10}$6s, 
4d$^{10}$7s,
4d$^{10}$5d, 
4d$^{10}$6d, 
4d$^{10}$7d, 
4d$^9$5s$^2$, 
4d$^9$5p$^2$,
4d$^9$5d$^2$, 
4d$^9$4f$^2$, 
4d$^9$5s5d, 
4d$^9$5s6d, 
4d$^9$5s6s,
4d$^9$4f5p, and 
4d$^9$4f6p (even parity) and 
4d$^{10}$5p, 
4d$^{10}$6p,
4d$^{10}$7p, 
4d$^{10}$4f, 
4d$^{10}$5f, 
4d$^{10}$6f, 
4d$^{10}$7f,
4d$^9$5s5p, 
4d$^9$5s6p, 
4d$^9$5s5f, 
4d$^9$5s6f, 
4d$^9$4f5s,
4d$^9$4f6s, 
4d$^9$4f5d, 
4d$^9$4f6d (odd parity). 
The core-polarization
parameters were those corresponding to a Rh-like \ion{Te}{viii} ionic core,
that is, $\alpha$$_d$ = 1.15\,a.u. 
\citep{fragaetal1976} 
and $r_\mathrm{c}$ $\equiv$
$<$$r$$>$$_\mathrm{4d}$ = 0.95\,a.u. The radial parameters of 
4d$^{10}$5s,
4d$^{10}$6s, 
4d$^{10}$5d, 
4d$^9$5p$^2$, 
4d$^{10}$5p, 
4d$^{10}$6p, and
4d$^9$5s5p 
configurations were optimized to minimize the differences
between the computed Hamiltonian eigenvalues and the experimental
energy levels published by 
\citet{crookerjoshi1964}, 
\citet{dunneosullivan1992}, and 
\citet{ryabtsevetal2007} giving rise to mean
deviations of 89\,m$^{-1}$ (even parity) and 13\,cm$^{-1}$ (odd
parity).

\paragraph{\ion{I}{vi}:\hspace{10mm}}
Thirty-two configurations were included in the HFR model used to
compute the atomic structure, i.e.,
5s$^2$, 
5p$^2$, 
5d$^2$,
4f$^2$, 
5f$^2$, 
5s5d, 
5s6d, 
5s7d, 
5s6s, 
5s7s, 
5p4f, 
5p5f, 
5p6f, 
5d6s,
5d7s, 
5d6d, and
5d7d for the even parity and 
5s5p, 
5s6p, 
5s7p, 
5s4f, 
5s5f,
5s6f, 
5s7f, 
5p5d, 
5p6d, 
5p7d, 
5p6s, 
5p7s, 
5d4f, 
5d5f, and
5d6f for the odd
parity. An ionic core of the type Pd-like \ion{I}{viii} was considered to
estimate the core-polarization effects with the parameters
$\alpha$$_d$ = 1.03\,a.u\@. 
\citep{johnsonetal1983} 
and $r_\mathrm{c}$ $\equiv$
$<$$r$$>$$_\mathrm{4d}$ = 0.90\,a.u. The semi-empirical optimization process
was carried out to adjust the radial parameters in 
5s$^2$, 
5p$^2$,
5s5d, 
5s6s, 
5s7s, 
5s5p, 
5s6p, 
5p5d, and 
5p6s 
with the experimental
energy levels taken from 
\citet{tauheedetal1997} 
giving rise to average
deviations of 72\,cm$^{-1}$ and 175\,cm$^{-1}$ for even and odd
parities, respectively.

\paragraph{}
The parameters adopted in our computations are summarized in 
Tables \ref{tab:sev:para} - \ref{tab:ivi:para}
while calculated and available experimental energies are compared in 
Tables \ref{tab:sev:ener} - \ref{tab:ivi:ener}, for 
\ion{Se}{v}, \ion{Sr}{iv-vii},\ion{Te}{vi}, and \ion{I}{vi}, respectively. 
Tables \ref{tab:sev:loggf} - \ref{tab:ivi:loggf} 
give the newly computed weighted oscillator strengths ($\log g_\mathrm{i}f_\mathrm{ik}$,
$_\mathrm{i}$ and $_\mathrm{k}$ are the indexes of the lower and upper energy level, respectively)
and transition probabilities ($g_\mathrm{k}A_\mathrm{ki}$, in s$^{-1}$) together with the
numerical values (in cm$^{-1}$) of the lower and upper energy levels
and the corresponding wavelengths (in \AA). In the final column of each
table, we also give the cancellation factor, $CF$, as defined by \citet{cowan1981}.
We note that very low values of this factor (typically $< 0.05$) indicate strong 
cancellation effects in the calculation of line
strengths. In these cases, the corresponding $\log g_\mathrm{i}f_\mathrm{ik}$ and $g_\mathrm{k}A_\mathrm{ki}$ values
could be very inaccurate and therefore need to be considered with some
care. 
Figure\,\ref{fig:normloggf} shows the newly calculated $\log g_\mathrm{i}f_\mathrm{ik}$ values 
from the X-ray to the far infrared wavelength range.

\paragraph{}
Radiative decay rates for some transitions in the same ions as those considered in the present 
work were reported in previous papers. More precisely, for \ion{Se}{v}, large-scale calculations for 
the 
4s$^2$ – 4s4p 
transitions were performed by \citet{liuetal2006} using the multiconfiguration 
Dirac–Fock (MCDF) method and by \citet{chencheng2010} using B-spline basis functions while the 
Relativistic Many Body Perturbation Theory (RMBPT), including the Breit interaction was used 
by \citet{safronovasafronova2010} to compute oscillator strengths for transitions between even-parity 
4s$^2$, 4p$^2$, 4s4d, 4d$^2$, 4p4f, 4f$^2$ 
and odd-parity 
4s4p, 4s4f, 4p4d, 4d4f 
states. In \ion{Sr}{iv}, transition 
probabilities and oscillator strengths for the electric dipole transitions involving the 
4s$^2$4p$^5$, 4s$^2$4p$^4$4d and 4s4p$^6$
configurations were obtained using the multiconfiguration Dirac–Fock approach by \citet{singhetal2013} 
and by \citet{aggarwalkeenan2014}. These works were subsequently extended by \citet{aggarwalkeenan2015} 
to transitions involving the 
4s$^2$4p$^5$, 4s$^2$4p$^4$4$\ell$, 4s4p$^6$, 4s$^2$4p$^4$5$\ell$, 4s$^2$4p$^3$4d$^2$, 4s4p$^5$4$\ell$, and 4s4p$^5$5$\ell$
configurations. For \ion{Sr}{vi}, relativistic quantum defect orbital (RQDO) and MCDF calculations of oscillator strengths were carried out by \citet{charromartin1998,charromartin2005} 
for the 
4p$^3$ – 4p$^2$5s 
transition array while the same methods were used by \citet{charromartin2002,charromartin2005} for investigating the 
4p$^2$ – 4p5s 
transitions in \ion{Sr}{vii}. In the case of \ion{Te}{vi}, \citet{choujohnson1997} performed third-order relativistic 
many-body perturbation theory (MBPT) calculations to evaluate the rates for 
5s – 5p 
transitions while 
\citet{migdalekgarmulewicz2000} used a relativistic ab initio model potential approach with explicit 
local exchange to produce oscillator strengths. In the same ion, the 
5s – 5p 
transition rates were 
also computed by \citet{glowackimigdalek2009} who employed a configuration-interaction method with numerical 
Dirac-Fock wave functions generated with noninteger outermost core shell occupation number while transition 
probabilities for 
5s-5p, 5p-5d, 4f-5d, and 5d-5f 
transitions were calculated by \citet{ivanova2011}. Finally, 
for \ion{I}{vi}, the oscillator strengths of the allowed and 
spin-forbidden 5s$^2$ $^{1}$S$_0$ – 5s5p $^{1,3}$P$_1$ transitions were 
evaluated by \citet{biemontetal2000} using the relativistic Hartree-Fock approach, including a 
core-polarization potential, and the MCDF method, as well as by 
\citet{glowackimigdalek2003} who used a relativistic configuration-interaction  method with numerical Dirac–Fock 
wavefunctions generated with an ab initio model potential allowing for core-valence correlation.

In order to estimate the overall reliability of the new atomic data obtained in the present work, 
we have compared them with some of the most recent and the most extensive calculations available in literature, 
selected among those listed hereabove. More particularly, in \ion{Se}{v}, we noticed that our oscillator strengths 
were in excellent agreement (within a few percent) with the RMBPT values published by \citet{safronovasafronova2010}. 
In the case of \ion{Sr}{iv}, we found a general agreement of about $20-30$\,\% between our results and the oscillator strengths 
published by \citet{aggarwalkeenan2015}, this agreement reaching even 10\,\% for the most intense lines. For \ion{Te}{vi}, 
the mean ratio between our transition probabilities and the few values reported by \citet{ivanova2011} was found to be 
equal to 1.18 while, for \ion{I}{vi}, a very good agreement (within 10\,\%) was observed when comparing the gf-values 
obtained in the present work with those computed by \citet{biemontetal2000} using either a relativistic Hartree-Fock 
or an MCDF model, taking core-valence correlation effects into account. All these comparisons 
allowed us to conclude that the accuracy of the new atomic data listed in the present paper should be about 20\,\%, 
at least for the strongest lines.

\section{Results}
\label{sect:results}

\begin{figure*}
   \resizebox{\hsize}{!}{\includegraphics{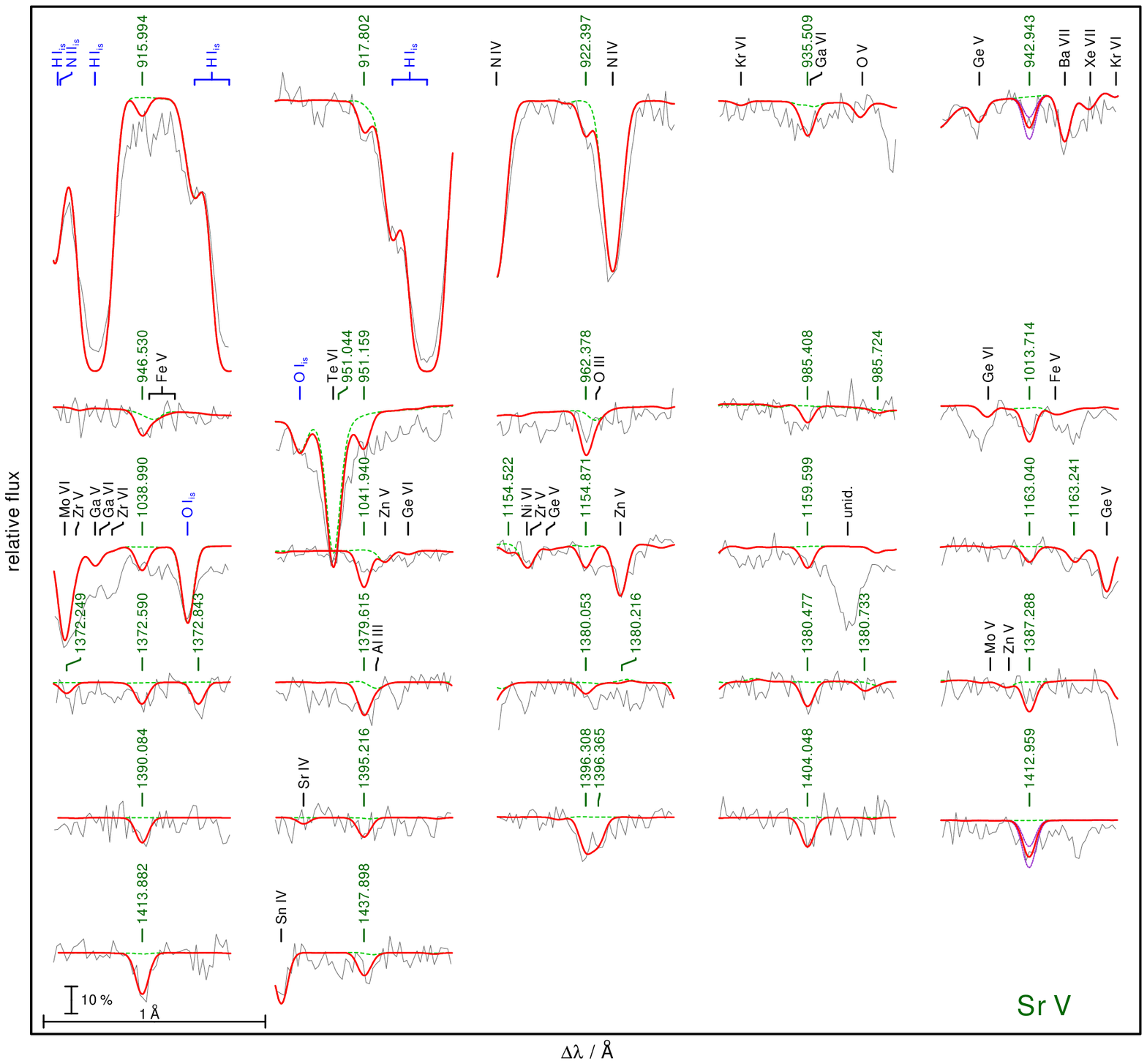}}
    \caption{ \ion{Sr}{v} lines in the observation (gray line) of \re, 
             labeled with their wavelengths from Table\,\ref{tab:srv:loggf}.
             The thick, red spectrum is calculated from our best model with a Sr mass fraction
             of $6.5 \times 10^{-4}$.
             The dashed, green line show a synthetic spectrum calculated without Sr. 
             In cases of \ionw{Sr}{v}{942.943} and
             \ionw{Sr}{v}{1412.959}, the red, dashed lines show two synthetic spectra
             calculated with Sr abundances that were increased and decreased by 0.3\,dex.
             The vertical bar indicates 10\,\% of the continuum flux.
             Identified lines are marked. ``is'' denotes interstellar.
            }
   \label{fig:sr}
\end{figure*}

In the FUSE and HST/STIS observations of \re, we newly identified 23 \ion{Sr}{v} lines,
listed in Table\,\ref{tab:srlineids} which complements Table\,A.1 of
\citet{hoyeretal2017}. Many more weak \ion{Sr}{v} are visible in our model spectra that are
not detectable in the noise of the available observations.
The models show that the strongest \ion{Sr}{vi} lines are located in the extreme ultraviolet (EUV) and
X-ray wavelength range while \ion{Sr}{iv} lines are too weak in general and fade within the noise of the 
available observations.
The observed \ion{Sr}{v} lines are well reproduced by our
model calculated with a mass fraction of $6.5 \times 10^{-4}$ (Fig.\,\ref{fig:sr}).
To estimate the abundance uncertainty from the error propagation of \Teff and \logg,
we evaluated models at the error limits (\Teffw{70\,000 \pm 2000}, \loggw{7.5 \pm 0.1}) and found that
it is smaller than 0.1\,dex. To consider the abundance uncertainties of other metals and the impact of
their background opacities, we finally adopted a Sr mass fraction of $6.5 \times 10^{-4}$ with an
uncertainty of 0.2\,dex (Fig.\,\ref{fig:sr} shows exemplarily the abundance dependence of two lines
in a $\left[-0.3\,\mathrm{dex},+0.3\,\mathrm{dex}\right]$ abundance interval). The determined Sr abundance 
matches well the abundance pattern of trans-iron elements in \re (Fig.\,\ref{fig:X}).
Their extreme overabundances are the result of efficient radiative levitation \citep{rauchetal2016mo}.

\begin{table}\centering
\caption{Identified \ion{Se}{v}, \ion{Sr}{v}, \ion{Te}{vi}, and \ion{I}{vi} lines in the UV spectrum of \re. 
         The wavelengths correspond to those in Table \ref{tab:srv:loggf}.}
\label{tab:srlineids}
\begin{tabular}{r@{\,}lr@{.}ll}
\hline\hline
\noalign{\smallskip}
\multicolumn{2}{c}{Ion} & \multicolumn{2}{c}{Wavelength\,/\,\AA} & Comment                        \\
\hline                                                                                            
Sr&{\sc v} &  915&994                               & uncertain                                   \\
Sr&{\sc v} &  917&802                               &                                             \\
 I&{\sc vi}&  919&210                               & blend, uncertain                            \\
 I&{\sc vi}&  919&555                               & blend, uncertain                            \\
Sr&{\sc v} &  922&397                               &                                             \\
Sr&{\sc v} &  935&509                               & blend, weak \ion{Ga}{vi}                    \\
Sr&{\sc v} &  942&943                               &                                             \\
Sr&{\sc v} &  946&530                               &                                             \\
Te&{\sc vi}&  951&021\tablefootmark{a}              &                                             \\
Sr&{\sc v} &  951&044                               & blend, strong \ion{Te}{vi}                  \\
Sr&{\sc v} &  951&159                               &                                             \\
Sr&{\sc v} &  962&378                               & blend, weak \ion{O}{iii}                    \\
Sr&{\sc v} &  985&408                               & uncertain                                   \\
Sr&{\sc v} & 1013&714                               &                                             \\
Sr&{\sc v} & 1038&990                               &                                             \\
Sr&{\sc v} & 1041&940                               &                                             \\
 I&{\sc vi}& 1053&389                               & weak                                        \\
 I&{\sc vi}& 1057&530                               & blend, strong \ion{Zn}{v}                   \\
Te&{\sc vi}& 1071&414\tablefootmark{a}              &                                             \\
Se&{\sc v} & 1094&691\tablefootmark{a}              &                                             \\
 I&{\sc vi}& 1120&301\tablefootmark{a}              &                                             \\
Se&{\sc v} & 1150&986\tablefootmark{a}              & shifted\tablefootmark{b} to 1151.016\,\AA   \\
 I&{\sc vi}& 1153&262                               &                                             \\
Sr&{\sc v} & 1154&871                               &                                             \\
Sr&{\sc v} & 1159&599                               & uncertain                                   \\
Sr&{\sc v} & 1163&040                               &                                             \\
Se&{\sc v} & 1227&446                               & shifted\tablefootmark{c} to 1227.540\,\AA   \\
Te&{\sc vi}& 1313&874                               &                                             \\
Sr&{\sc v} & 1372&590                               &                                             \\
Sr&{\sc v} & 1372&843                               &                                             \\
Sr&{\sc v} & 1379&615                               & blend, weak \ion{Al}{iii}                   \\
Sr&{\sc v} & 1380&053                               & uncertain                                   \\
Sr&{\sc v} & 1380&477                               &                                             \\
Sr&{\sc v} & 1387&288                               & uncertain                                   \\
Sr&{\sc v} & 1390&084                               &                                             \\
Sr&{\sc v} & 1395&216                               & uncertain                                   \\
Sr&{\sc v} & 1396&308                               & blend, \ionw{Sr}{v}{1396.365}               \\
Sr&{\sc v} & 1396&365                               & blend, \ionw{Sr}{v}{1396.308}               \\
Sr&{\sc v} & 1404&048                               &                                             \\
Sr&{\sc v} & 1412&958                               &                                             \\
Sr&{\sc v} & 1413&882                               &                                             \\
Sr&{\sc v} & 1437&898                               &                                             \\
Se&{\sc v} & 1451&779                               & shifted\tablefootmark{d} to 1451.653\,\AA   \\
Se&{\sc v} & 1454&292                               &                                             \\
\hline
\end{tabular}
\tablefoot{
\tablefoottext{a}{Identified by \citet{werneretal2012},
\tablefoottext{b}{Shifted to match observation. \citet{raobadami1931} measured 1151.96\,\AA.},
\tablefoottext{c}{Shifted to match observation. \citet{raobadami1931} measured 1227.58\,\AA.},
\tablefoottext{d}{Shifted to match observation.}
}
}
\end{table}

\begin{figure}
   \resizebox{\hsize}{!}{\includegraphics{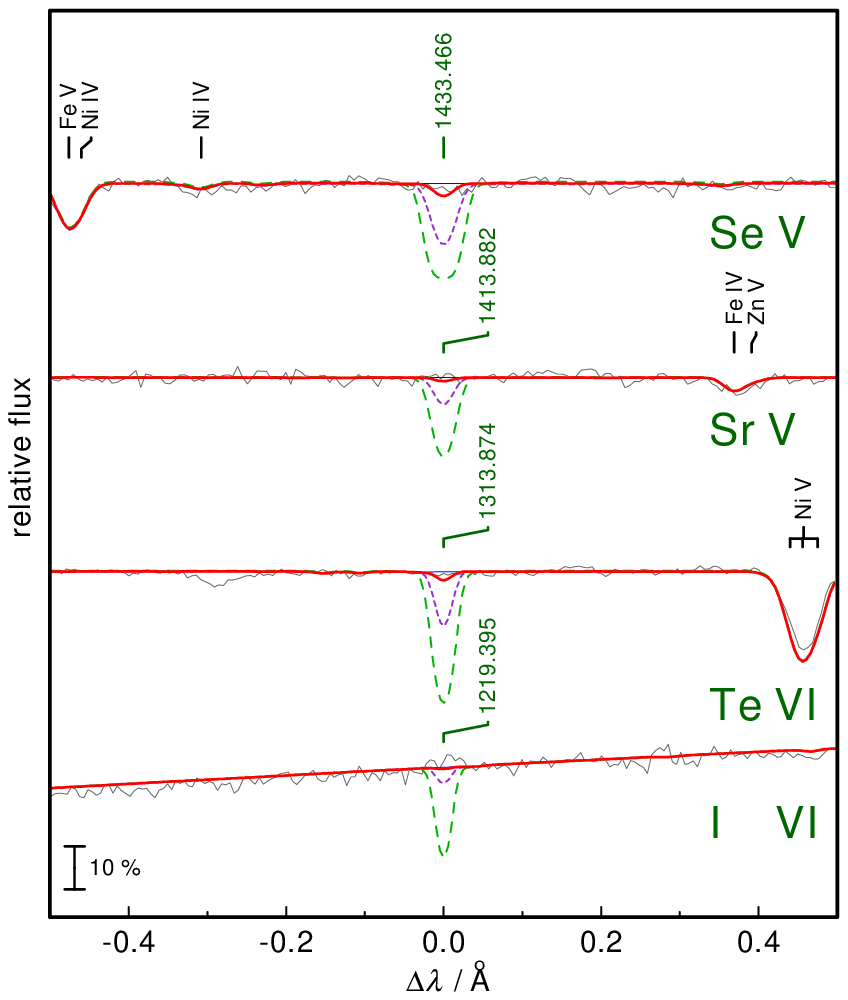}}
    \caption{STIS observation of \gb (gray) compared with synthetic line profiles of
             \ionw{Se}{v}{1433.466},
             \ionw{Sr}{v}{1413.882}, 
             \ionw{Te}{vi}{1313.874}, and
             \ionw{I}{vi}{1219.395}.
             The models were calculated with four abundances of the respective elements,
             without (thin, blue),
             with   100 times (thick, red), 
                   1000 times (short dashed, violet) and
                  10000 times solar abundance (long dashed, green).
            }
   \label{fig:srgb}
\end{figure}

\begin{figure*}
   \resizebox{\hsize}{!}{\includegraphics{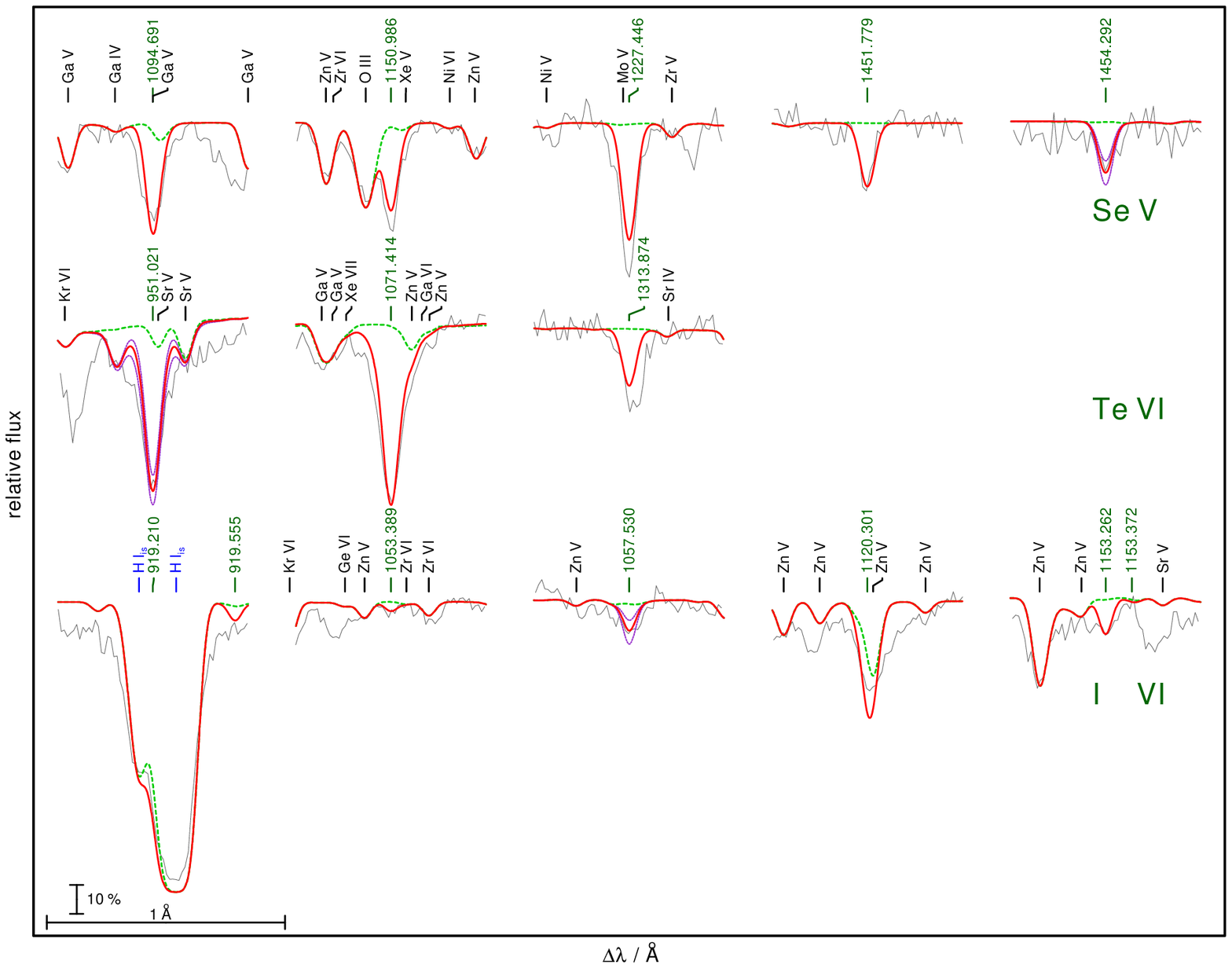}}
    \caption{As Fig.\,\ref{fig:sr}, for 
             \ion{Se}{v} (top, mass fraction of $1.6 \pm 1.0 \times 10^{-3}$ in the model), 
             \ion{Te}{vi} (middle, $2.5 \times 10^{-4}$), and 
             \ion{I}{vi} (bottom, $1.4 \times 10^{-5}$)
              lines.
              Abundance dependencies ($\pm 0.3\,\mathrm{dex}$, red, dashed lines) are demonstrated for
              \ionw{Se}{v}{1454.292},
              \ionw{Te}{vi}{951.021}, and
              \ionw{I}{vi}{1057.530}.
            }
    \label{fig:setei}
\end{figure*}

Five \ion{Se}{v}, three \ion{Te}{vi}, and four \ion{I}{vi} lines are used for the
abundance determination of these elements (Fig.\,\ref{fig:setei}).
We measured mass fractions of 
$1.6\times 10^{-3}$, 
$2.5\times 10^{-4}$, and 
$1.4\times 10^{-5}$
for Se, Te, and I, respectively.
These agree well with the expectations from the abundance pattern of trans-iron
elements in \re (Fig.\,\ref{fig:X}).

A very weak impact of Se, Sr, Te, and I lines is noticeable in the EUV wavelength range.
The so-called EUV problem, that is, the discrepancy between model and observation in this
wavelength range \citep[cf.,][]{hoyeretal2017} is, however, not significantly reduced.
We note that \citet{prevaletal2017} showed recently
that improved, larger photoionization cross-section of Ni can reduce this discrepancy.

The search for Se, Sr, Te, and I lines in the FUSE and HST/STIS observations of \gb was entirely negative.
Figure\,\ref{fig:srgb} shows the most prominent lines that are predicted by our models, namely
\ionw{Se}{v}{1433.466} ($\log g_\mathrm{i}f_\mathrm{ik}$ value of 0.16),
\ionw{Sr}{v}{1413.882} (0.82), 
\ionw{Te}{vi}{1313.874} ($-0.05$), and
\ionw{I}{vi}{1219.395} ($-0.63$).
For all these elements, the upper abundance limit is $\approx 100$ times the solar abundance.

\begin{acknowledgements}
TR and DH are supported by the German Aerospace Center (DLR, grants 05\,OR\,1507 and 50\,OR\,1501, respectively).
The GAVO project had been supported by the Federal Ministry of Education and Research (BMBF) 
at T\"ubingen (05\,AC\,6\,VTB, 05\,AC\,11\,VTB) and is funded
at Heidelberg (05\,AC\,11\,VH3).
Financial support from the Belgian FRS-FNRS is also acknowledged. 
PQ is research director of this organization.
Some of the data presented in this paper were obtained from the
Mikulski Archive for Space Telescopes (MAST). STScI is operated by the
Association of Universities for Research in Astronomy, Inc., under NASA
contract NAS5-26555. Support for MAST for non-HST data is provided by
the NASA Office of Space Science via grant NNX09AF08G and by other
grants and contracts. 
The TIRO (\url{http://astro-uni-tuebingen.de/~TIRO}),
    TMAD (\url{http://astro-uni-tuebingen.de/~TMAD}), and 
    TOSS (\url{http://astro-uni-tuebingen.de/~TOSS}) services
were constructed as part of the T\"ubingen project of the German Astrophysical Virtual Observatory
(GAVO, \url{http://www.g-vo.org}).
This research has made use of 
NASA's Astrophysics Data System and
the SIMBAD database, operated at CDS, Strasbourg, France.
\end{acknowledgements}

\bibliographystyle{aa}
\bibliography{30383}

\begin{appendix}
\onecolumn

\section{Additional tables}
\label{app:addtabs}

\begin{longtable}{llrrrl}
\caption{\label{tab:sev:para}Radial parameters (in cm$^{-1}$) adopted for the calculations in \ion{Se}{v}.} \\
\hline\hline
\noalign{\smallskip}
Configuration & Parameter & HFR & Fitted & Ratio & Note$^a$ \\ 
\noalign{\smallskip}
\hline
\endfirsthead
\caption{continued.}\\
\hline\hline
\noalign{\smallskip}
Configuration & Parameter & HFR & Fitted & Ratio & Note$^a$ \\
\noalign{\smallskip}
\hline
\noalign{\smallskip}
\endhead
\hline
\noalign{\smallskip}
\endfoot
\noalign{\smallskip}
\multicolumn{6}{c}{Even parity} \\
4s$^2$           & E$_{av}$          &    7245 &   7181 &       &    \\
4p$^2$           & E$_{av}$          &  220372 & 224026 &       &    \\
                 & F$^2$(4p,4p)      &   59154 &  53762 & 0.909 &    \\
                 & $\alpha$          &       0 &    457 &       &    \\
                 & $\zeta$$_{4p}$    &    3088 &   3450 & 1.117 &    \\
4s4d             & E$_{av}$          &  258748 & 261651 &       &    \\
                 & $\zeta$$_{4d}$    &     182 &    226 & 1.243 &    \\
\hline
\end{longtable}
$^{(a)}$ F: Fixed parameter value; Rn: ratios of these parameters have been fixed in the fitting process.\\
\noindent

\begin{longtable}{llrrrl}
\caption{\label{tab:sriv:para}Radial parameters (in cm$^{-1}$) adopted for the calculations in \ion{Sr}{iv}.} \\
\hline\hline
\noalign{\smallskip}
Configuration & Parameter & HFR & Fitted & Ratio & Note$^a$ \\ 
\noalign{\smallskip}
\hline
\endfirsthead
\caption{continued.}\\
\hline\hline
\noalign{\smallskip}
Configuration & Parameter & HFR & Fitted & Ratio & Note$^a$ \\
\noalign{\smallskip}
\hline
\noalign{\smallskip}
\endhead
\hline
\noalign{\smallskip}
\endfoot
\noalign{\smallskip}
\multicolumn{6}{c}{Odd parity} \\
4p$^5$           & E$_{av}$          &   17382 &  17802 &       &    \\
                 & $\zeta$$_{4p}$    &    6193 &   6602 & 1.066 &    \\
4p$^4$5p         & E$_{av}$          &  297016 & 284223 &       &    \\
                 & F$^2$(4p,4p)      &   72698 &  59785 & 0.822 &    \\
                 & $\alpha$          &       0 &  $-$75 &       &    \\
                 & $\zeta$$_{4p}$    &    6661 &   6788 & 1.019 &    \\
                 & $\zeta$$_{5p}$    &    1144 &   1371 & 1.198 &    \\
\hline
\end{longtable}
$^{(a)}$ F: Fixed parameter value; Rn: ratios of these parameters have been fixed in the fitting process.\\
\noindent

\begin{longtable}{llrrrl}
\caption{\label{tab:srv:para}Radial parameters (in cm$^{-1}$) adopted for the calculations in \ion{Sr}{v}.} \\
\hline\hline
\noalign{\smallskip}
Configuration & Parameter & HFR & Fitted & Ratio &  \\ 
\noalign{\smallskip}
\hline
\endfirsthead
\caption{continued.}\\
\hline\hline
\noalign{\smallskip}
Configuration & Parameter & HFR & Fitted & Ratio & \\
\noalign{\smallskip}
\hline
\noalign{\smallskip}
\endhead
\hline
\noalign{\smallskip}
\endfoot
\noalign{\smallskip}
\multicolumn{5}{c}{Even parity} \\
4p$^4$           & E$_{av}$          &   26722 &  27025 &       \\
                 & F$^2$(4p,4p)      &   73003 &  66387 & 0.909 \\
                 & $\alpha$          &       0 & $-$116 &       \\
                 & $\zeta$$_{4p}$    &    6711 &   7128 & 1.062 \\
4p$^3$5p         & E$_{av}$          &  356017 & 342959 &       \\
                 & F$^2$(4p,4p)      &   75078 &  60470 & 0.805 \\
                 & $\alpha$          &       0 &  $-$50 &       \\
\hline
\end{longtable}
\noindent

\begin{longtable}{llrrrl}
\caption{\label{tab:srvi:para}Radial parameters (in cm$^{-1}$) adopted for the calculations in \ion{Sr}{vi}.} \\
\hline\hline
\noalign{\smallskip}
Configuration & Parameter & HFR & Fitted & Ratio & \\ 
\noalign{\smallskip}
\hline
\endfirsthead
\caption{continued.}\\
\hline\hline
\noalign{\smallskip}
Configuration & Parameter & HFR & Fitted & Ratio & \\
\noalign{\smallskip}
\hline
\noalign{\smallskip}
\endhead
\hline
\noalign{\smallskip}
\endfoot
\noalign{\smallskip}
\multicolumn{5}{c}{Odd parity} \\
4p$^3$           & E$_{av}$          &   38168 &  38218 &       \\
                 & F$^2$(4p,4p)      &   75420 &  67860 & 0.900 \\
                 & $\alpha$          &       0 &  $-$58 &       \\
                 & $\zeta$$_{4p}$    &    7245 &   7847 & 1.083 \\[0.3cm]
\multicolumn{5}{c}{Even parity} \\
4s4p$^4$         & E$_{av}$          &  215385 & 207203 &       \\
                 & F$^2$(4p,4p)      &   75398 &  65586 & 0.870 \\
                 & $\alpha$          &       0 &    360 &       \\
                 & $\zeta$$_{4p}$    &    7243 &   7548 & 1.042 \\
                 & G$^1$(4s,4p)      &  100681 &  84856 & 0.843 \\
4p$^2$5s         & E$_{av}$          &  361390 & 351438 &       \\
                 & F$^2$(4p,4p)      &   76922 &  70839 & 0.921 \\
                 & $\alpha$          &       0 & $-$658 &       \\
                 & $\zeta$$_{4p}$    &    7644 &   8279 & 1.083 \\
                 & G$^1$(4p,5s)      &    8511 &   6300 & 0.740 \\
\hline
\end{longtable}
\noindent

\begin{longtable}{llrrrl}
\caption{\label{tab:srvii:para}Radial parameters (in cm$^{-1}$) adopted for the calculations in \ion{Sr}{vii}.} \\
\hline\hline
\noalign{\smallskip}
Configuration & Parameter & HFR & Fitted & Ratio & \\ 
\noalign{\smallskip}
\hline
\endfirsthead
\caption{continued.}\\
\hline\hline
\noalign{\smallskip}
Configuration & Parameter & HFR & Fitted & Ratio & \\
\noalign{\smallskip}
\hline
\noalign{\smallskip}
\endhead
\hline
\noalign{\smallskip}
\endfoot
\noalign{\smallskip}
\multicolumn{5}{c}{Even parity} \\
4p$^2$           & E$_{av}$          &   30564 &  31486 &       \\
                 & F$^2$(4p,4p)      &   77689 &  70341 & 0.905 \\
                 & $\alpha$          &       0 &  $-$11 &       \\
                 & $\zeta$$_{4p}$    &    7791 &   8364 & 1.073 \\[0.3cm]
\multicolumn{5}{c}{Odd parity} \\
4s4p$^3$         & E$_{av}$          &  202602 & 200219 &       \\
                 & F$^2$(4p,4p)      &   77662 &  64903 & 0.836 \\
                 & $\alpha$          &       0 &    401 &       \\
                 & $\zeta$$_{4p}$    &    7783 &   8251 & 1.060 \\
                 & G$^1$(4s,4p)      &  103310 &  89586 & 0.867 \\
4p5s             & E$_{av}$          &  396223 & 391256 &       \\
                 & $\zeta$$_{4p}$    &    8187 &   8710 & 1.064 \\
                 & G$^1$(4p,5s)      &    9207 &   6618 & 0.719 \\
\hline
\end{longtable}
\noindent

\begin{longtable}{llrrrl}
\caption{\label{tab:tevi:para}Radial parameters (in cm$^{-1}$) adopted for the calculations in \ion{Te}{vi}.} \\
\hline\hline
\noalign{\smallskip}
Configuration & Parameter & HFR & Fitted & Ratio & \\ 
\noalign{\smallskip}
\hline
\endfirsthead
\caption{continued.}\\
\hline\hline
\noalign{\smallskip}
Configuration & Parameter & HFR & Fitted & Ratio & \\
\noalign{\smallskip}
\hline
\noalign{\smallskip}
\endhead
\hline
\noalign{\smallskip}
\endfoot
\noalign{\smallskip}
\multicolumn{6}{c}{Even parity} \\
4d$^{10}$5s      & E$_{av}$          &    8200 &   8203 &       &    \\
4d$^{10}$6s      & E$_{av}$          &  281972 & 279957 &       &    \\
4d$^{10}$5d      & E$_{av}$          &  242153 & 242724 &       &    \\
                 & $\zeta$$_{5d}$    &     544 &    651 & 1.198 &    \\
4d$^9$5p$^2$     & E$_{av}$          &  514322 & 508542 &       &    \\
                 & F$^2$(5p,5p)      &   54096 &  55453 & 1.025 &    \\
                 & $\zeta$$_{4d}$    &    4672 &   4912 & 1.051 &    \\
\hline
\end{longtable}
$^{(a)}$ F: Fixed parameter value; Rn: ratios of these parameters have been fixed in the fitting process.\\
\noindent

\begin{longtable}{llrrrl}
\caption{\label{tab:ivi:para}Radial parameters (in cm$^{-1}$) adopted for the calculations in \ion{I}{vi}.} \\
\hline\hline
\noalign{\smallskip}
Configuration & Parameter & HFR & Fitted & Ratio & \\ 
\noalign{\smallskip}
\hline
\endfirsthead
\caption{continued.}\\
\hline\hline
\noalign{\smallskip}
Configuration & Parameter & HFR & Fitted & Ratio & \\
\noalign{\smallskip}
\hline
\noalign{\smallskip}
\endhead
\hline
\noalign{\smallskip}
\endfoot
\noalign{\smallskip}
\multicolumn{6}{c}{Even parity} \\
5s$^2$           & E$_{av}$          &    6244 &   6194 &       &    \\
5p$^2$           & E$_{av}$          &  216536 & 221549 &       &    \\
                 & F$^2$(5p,5p)      &   55129 &  50155 & 0.910 &    \\
                 & $\alpha$          &       0 &    921 &       &    \\
                 & $\zeta$$_{5p}$    &    8475 &   9377 & 1.106 &    \\
5s5d             & E$_{av}$          &  252535 & 256150 &       &    \\
                 & $\zeta$$_{5d}$    &     628 &    802 & 1.277 &    \\
\hline
\end{longtable}
$^{(a)}$ F: Fixed parameter value; Rn: ratios of these parameters have been fixed in the fitting process.\\
\noindent

\begin{longtable}{rrrrl}
\caption{\label{tab:sev:ener} Comparison between available experimental and calculated energy levels in \ion{Se}{v}. Energies are given in cm$^{-1}$.} \\
\hline\hline
\noalign{\smallskip}
E$_{exp}$$^a$  & E$_{calc}$$^b$ & $\Delta$E & J & Leading components (in \%) in $LS$ coupling$^c$ \\ 
\noalign{\smallskip}
\hline
\endfirsthead
\caption{continued.}\\
\hline\hline
\noalign{\smallskip}
E$_{exp}$$^a$  & E$_{calc}$$^b$ & $\Delta$E & J & Leading components (in \%) in $LS$ coupling$^c$ \\
\noalign{\smallskip}
\hline
\noalign{\smallskip}
\endhead
\hline
\noalign{\smallskip}
\endfoot
\noalign{\smallskip}
\multicolumn{5}{c}{Even parity} \\
        0 &        0 &      0 & 0 &   97  4s$^2$ $^1$S \\  
   211794 &   211776 &     18 & 0 &   98  4p$^2$ $^3$P \\  
   213203 &   213206 &   $-$3 & 2 &   63  4p$^2$ $^1$D  +   20  4p$^2$ $^3$P  +   17  4s4d $^1$D \\ 
   214091 &   214117 &  $-$26 & 1 &  100  4p$^2$ $^3$P \\  
   218615 &   218604 &     11 & 2 &   80  4p$^2$ $^3$P  +   15  4p$^2$ $^1$D \\  
   248787 &   248787 &      0 & 0 &   94  4p$^2$ $^1$S \\  
   257536 &   257535 &      1 & 1 &  100  4s4d $^3$D \\
\hline
\end{longtable}
\noindent
$^{(a)}$ From \citet{churilovjoshi1995}.\\
$^{(b)}$ This work.\\
$^{(c)}$ Only the first three components larger than 5\% are given.

\begin{longtable}{rrrrl}
\caption{\label{tab:sriv:ener} Comparison between available experimental and calculated energy levels in \ion{Sr}{iv}. Energies are given in cm$^{-1}$.} \\
\hline\hline
\noalign{\smallskip}
E$_{exp}$$^a$  & E$_{calc}$$^b$ & $\Delta$E & J & Leading components (in \%) in $LS$ coupling$^c$ \\ 
\noalign{\smallskip}
\hline
\endfirsthead
\caption{continued.}\\
\hline\hline
\noalign{\smallskip}
E$_{exp}$$^a$  & E$_{calc}$$^b$ & $\Delta$E & J & Leading components (in \%) in $LS$ coupling$^c$ \\
\noalign{\smallskip}
\hline
\noalign{\smallskip}
\endhead
\hline
\noalign{\smallskip}
\endfoot
\noalign{\smallskip}
\multicolumn{5}{c}{Odd parity} \\

        0.00 &        0 &      0 & 1.5 &   97  4p$^5$ $^2$P \\     
     9727.90 &     9728 &      0 & 0.5 &   97  4p$^5$ $^2$P \\  
   267529.26 &   267364 &    165 & 1.5 &   68  4p$^4$($^3$P)5p $^4$P  +    8  4p$^4$($^3$P)5p $^4$S  +    7  4p$^4$($^3$P)5p $^2$P \\   
   267537.32 &   267389 &    148 & 2.5 &   79  4p$^4$($^3$P)5p $^4$P  +   16  4p$^4$($^3$P)5p $^4$D \\    
   270350.30 &   270269 &     81 & 0.5 &   54  4p$^4$($^3$P)5p $^4$P  +   19  4p$^4$($^3$P)5p $^2$P  +   14  4p$^4$($^1$D)5p $^2$P \\   
   271249.46 &   271591 & $-$342 & 3.5 &   93  4p$^4$($^3$P)5p $^4$D  +    7  4p$^4$($^1$D)5p $^2$F \\   
   271328.56 &   271425 &  $-$96 & 2.5 &   65  4p$^4$($^3$P)5p $^2$D  +   20  4p$^4$($^3$P)5p $^4$D  +    7  4p$^4$($^1$D)5p $^2$F \\
\hline
\end{longtable}
\noindent
$^{(a)}$ From \citet{sansonetti2012}.\\
$^{(b)}$ This work.\\
$^{(c)}$ Only the first three components larger than 5\% are given.

\begin{longtable}{rrrrl}
\caption{\label{tab:srv:ener} Comparison between available experimental and calculated energy levels in \ion{Sr}{v}. Energies are given in cm$^{-1}$.} \\
\hline\hline
\noalign{\smallskip}
E$_{exp}$$^a$  & E$_{calc}$$^b$ & $\Delta$E & J & Leading components (in \%) in $LS$ coupling$^c$ \\ 
\noalign{\smallskip}
\hline
\endfirsthead
\caption{continued.}\\
\hline\hline
\noalign{\smallskip}
E$_{exp}$$^a$  & E$_{calc}$$^b$ & $\Delta$E & J & Leading components (in \%) in $LS$ coupling$^c$ \\
\noalign{\smallskip}
\hline
\noalign{\smallskip}
\endhead
\hline
\noalign{\smallskip}
\endfoot
\noalign{\smallskip}
\multicolumn{5}{c}{Even parity} \\

        0.00 &    $-$15 &     15 & 2 &   91  4p$^4$ $^3$P           +    6  4p$^4$ $^1$D \\    
     8307.88 &     8276 &     32 & 1 &   97  4p$^4$ $^3$P \\   
     8718.47 &     8768 &  $-$50 & 0 &   89  4p$^4$ $^3$P           +    8  4p$^4$ $^1$S \\  
    20310.51 &    20311 &      0 & 2 &   91  4p$^4$ $^1$D           +    6  4p$^4$ $^3$P \\    
    44050.41 &    44048 &      2 & 0 &   88  4p$^4$ $^1$S           +    8  4p$^4$ $^3$P \\  
   313726.94 &   313649 &     78 & 1 &   89  4p$^3$($^4$S)5p $^5$P \\    
   314143.18 &   314071 &     72 & 2 &   82  4p$^3$($^4$S)5p $^5$P  +   10  4p$^3$($^4$S)5p $^3$P \\
\hline
\end{longtable}
\noindent
$^{(a)}$ From \citet{sansonetti2012}.\\
$^{(b)}$ This work.\\
$^{(c)}$ Only the first three components that are larger than 5\% are given.

\begin{longtable}{rrrrl}
\caption{\label{tab:srvi:ener} Comparison between available experimental and calculated energy levels in \ion{Sr}{vi}. Energies are given in cm$^{-1}$.} \\
\hline\hline
\noalign{\smallskip}
E$_{exp}$$^a$  & E$_{calc}$$^b$ & $\Delta$E & J & Leading components (in \%) in $LS$ coupling$^c$ \\ 
\noalign{\smallskip}
\hline
\endfirsthead
\caption{continued.}\\
\hline\hline
\noalign{\smallskip}
E$_{exp}$$^a$  & E$_{calc}$$^b$ & $\Delta$E & J & Leading components (in \%) in $LS$ coupling$^c$ \\
\noalign{\smallskip}
\hline
\noalign{\smallskip}
\endhead
\hline
\noalign{\smallskip}
\endfoot
\noalign{\smallskip}
\multicolumn{5}{c}{Odd parity} \\
        0.0  &       1 &   $-$1 & 1.5 &   93  4p$^3$ $^4$S \\    
    20134.7  &   20114 &     21 & 1.5 &   82  4p$^3$ $^2$D  +   13  4p$^3$ $^2$P \\    
    23527.0  &   23546 &  $-$19 & 2.5 &   97  4p$^3$ $^2$D \\    
    38531.1  &   38544 &  $-$13 & 0.5 &   97  4p$^3$ $^2$P \\  
    43566.9  &   43555 &     12 & 1.5 &   79  4p$^3$ $^2$P  +   15  4p$^3$ $^2$D \\    
\multicolumn{5}{c}{Even parity} \\
   153170.8  &  153175 &   $-$4 & 2.5 &   90  4s4p$^4$ $^4$P         +    8  4p$^2$($^3$P)4d $^4$P \\
\hline
\end{longtable}
\noindent
$^{(a)}$ From \citet{sansonetti2012}.\\
$^{(b)}$ This work.\\
$^{(c)}$ Only the first three components that are larger than 5\% are given.

\begin{longtable}{rrrrl}
\caption{\label{tab:srvii:ener} Comparison between available experimental and calculated energy levels in \ion{Sr}{vii}. Energies are given in cm$^{-1}$.} \\
\hline\hline
\noalign{\smallskip}
E$_{exp}$$^a$  & E$_{calc}$$^b$ & $\Delta$E & J & Leading components (in \%) in $LS$ coupling$^c$ \\ 
\noalign{\smallskip}
\hline
\endfirsthead
\caption{continued.}\\
\hline\hline
\noalign{\smallskip}
E$_{exp}$$^a$  & E$_{calc}$$^b$ & $\Delta$E & J & Leading components (in \%) in $LS$ coupling$^c$ \\
\noalign{\smallskip}
\hline
\noalign{\smallskip}
\endhead
\hline
\noalign{\smallskip}
\endfoot
\noalign{\smallskip}
\multicolumn{5}{c}{Even parity} \\
        0  &       0 &      0 & 0 &   92  4p$^2$ $^3$P   +    5  4p$^2$ $^1$S \\  
     6845  &    6845 &      0 & 1 &   98  4p$^2$ $^3$P  \\    
    12545  &   12545 &      0 & 2 &   82  4p$^2$ $^3$P   +   15  4p$^2$ $^1$D \\    
    28490  &   28490 &      0 & 2 &   82  4p$^2$ $^1$D   +   15  4p$^2$ $^3$P  \\    
    51814  &   51814 &      0 & 0 &   91  4p$^2$ $^1$S   +    5  4p$^2$ $^3$P  \\     
\multicolumn{5}{c}{Odd parity} \\
   167806  &  168195 & $-$389 & 1 &   85  4s4p$^3$ $^3$D   +    7  4p4d $^3$D       +    5  4s4p$^3$ $^3$P  \\
\hline
\end{longtable}
\noindent
$^{(a)}$ From \citet{sansonetti2012}.\\
$^{(b)}$ This work.\\
$^{(c)}$ Only the first three components that are larger than 5\% are given.

\begin{longtable}{rrrrl}
\caption{\label{tab:tevi:ener} Comparison between available experimental and calculated energy levels in \ion{Te}{vi}. Energies are given in cm$^{-1}$.} \\
\hline\hline
\noalign{\smallskip}
E$_{exp}$$^a$  & E$_{calc}$$^b$ & $\Delta$E & J & Leading components (in \%) in $LS$ coupling$^c$ \\ 
\noalign{\smallskip}
\hline
\endfirsthead
\caption{continued.}\\
\hline\hline
\noalign{\smallskip}
E$_{exp}$$^a$  & E$_{calc}$$^b$ & $\Delta$E & J & Leading components (in \%) in $LS$ coupling$^c$ \\
\noalign{\smallskip}
\hline
\noalign{\smallskip}
\endhead
\hline
\noalign{\smallskip}
\endfoot
\noalign{\smallskip}
\multicolumn{5}{c}{Even parity} \\
        0.0  &       0 &      0 & 0.5 &   99  4d$^{10}$5s $^2$S \\  
   238087.6  &  238088 &      0 & 1.5 &   99  4d$^{10}$5d $^2$D \\  
   239736.2  &  239736 &      0 & 2.5 &   99  4d$^{10}$5d $^2$D \\  
   278436.2  &  278436 &      0 & 0.5 &  100  4d$^{10}$6s $^2$S \\  
   480834.0  &  480935 & $-$101 & 2.5 &   49  4d$^9$5p$^2$ $^4$D  +   13  4d$^9$5p$^2$ $^4$F  +    9  4d$^9$5p$^2$ $^4$P \\ 
   488295.0  &  488184 &    111 & 1.5 &   64  4d$^9$5p$^2$ $^4$D  +   10  4d$^9$5p$^2$ $^2$D  +    9  4d$^9$5p$^2$ $^4$F \\ 
   489538.0  &  489652 & $-$114 & 0.5 &   56  4d$^9$5p$^2$ $^2$S  +   19  4d$^9$5p$^2$ $^2$P  +    8  4d$^9$5p$^2$ $^4$D \\
\hline
\end{longtable}
\noindent
$^{(a)}$ From \citet{crookerjoshi1964}, \citet{dunneosullivan1992}, and \citet{ryabtsevetal2007}.\\
$^{(b)}$ This work.\\
$^{(c)}$ Only the first three components larger than 5\% are given.

\begin{longtable}{rrrrl}
\caption{\label{tab:ivi:ener} Comparison between available experimental and calculated energy levels in \ion{I}{vi}. Energies are given in cm$^{-1}$.} \\
\hline\hline
\noalign{\smallskip}
E$_{exp}$$^a$  & E$_{calc}$$^b$ & $\Delta$E & J & Leading components (in \%) in $LS$ coupling$^c$ \\ 
\noalign{\smallskip}
\hline
\endfirsthead
\caption{continued.}\\
\hline\hline
\noalign{\smallskip}
E$_{exp}$$^a$  & E$_{calc}$$^b$ & $\Delta$E & J & Leading components (in \%) in $LS$ coupling$^c$ \\
\noalign{\smallskip}
\hline
\noalign{\smallskip}
\endhead
\hline
\noalign{\smallskip}
\endfoot
\noalign{\smallskip}
\multicolumn{5}{c}{Even parity} \\
        0.0 &        0 &      0 & 0 &   98  5s$^2$ $^1$S \\  
   200084.8 &   199954 &    131 & 0 &   91  5p$^2$ $^3$P  +   9  5p$^2$ $^1$S \\  
   208474.5 &   208663 & $-$188 & 1 &  100  5p$^2$ $^3$P \\   
   209431.7 &   209483 &  $-$51 & 2 &   54  5p$^2$ $^1$D  +  32  5p$^2$ $^3$P  +  13  5s5d $^1$D \\ 
   221983.8 &   221854 &    130 & 2 &   67  5p$^2$ $^3$P  +  23  5p$^2$ $^1$D  +  10  5s5d $^1$D \\ 
   245658.5 &   245671 &  $-$12 & 0 &   87  5p$^2$ $^1$S  +   9  5p$^2$ $^3$P \\   
   251816.9 &   251813 &      4 & 1 &   99  5s5d   $^3$D \\
\hline
\end{longtable}
\noindent
$^{(a)}$ From \citet{tauheedetal1997}.\\
$^{(b)}$ This work.\\
$^{(c)}$ Only the first three components larger than 5\% are given.

\begin{longtable}{rrccrrccrrr}
\caption{\label{tab:sev:loggf}Calculated HFR oscillator strengths ($\log g_\mathrm{i}f_\mathrm{ik}$) and transition probabilities ($g_\mathrm{k}A_\mathrm{ki}$) 
                               in \ion{Se}{v}. CF is the
                               absolute value of the cancellation factor as defined by \citet{cowan1981}. 
                               In Cols\@. 3 and 6, e represents even and
o odd.} \\
\hline\hline
\noalign{\smallskip}
                 & \multicolumn{3}{c}{Lower level} && \multicolumn{3}{c}{Upper level} &           &      &    \\
\cline{2-4}\cline{6-8}
\noalign{\smallskip}
Wavelength{$^a$} / \AA & Energy{$^b$} / cm$^{-1}$ & Parity & $J$  && Energy{$^b$} / cm$^{-1}$ & Parity & $J$  & $\log g_\mathrm{i}f_\mathrm{ik}$ & $g_\mathrm{k}A_\mathrm{ki}$ / s$^{-1}$  & CF \\
\noalign{\smallskip}
\hline
\endfirsthead
\caption{continued.}\\
\hline\hline
\noalign{\smallskip}
                 & \multicolumn{3}{c}{Lower level} && \multicolumn{3}{c}{Upper level} &           &      &    \\
\cline{2-4}\cline{6-8}
\noalign{\smallskip}
Wavelength{$^a$} / \AA & Energy{$^b$} / cm$^{-1}$ & Parity & $J$  && Energy{$^b$} / cm$^{-1}$ & Parity & $J$  & $\log g_\mathrm{i}f_\mathrm{ik}$ & $g_\mathrm{k}A_\mathrm{ki}$ / s$^{-1}$  & CF \\
\noalign{\smallskip}
\hline
\noalign{\smallskip}
\endhead
\hline
\noalign{\smallskip}
\endfoot
\noalign{\smallskip}
\hline
\noalign{\smallskip}
\multicolumn{11}{p{15.4cm}}{{Notes.\,
                             $^{(a)}$\,All wavelengths (given in vacuum for                         $\lambda <   2000\,\mathrm{\AA}$, 
                                                                   air for  $2000\,\mathrm{\AA} \le \lambda \le 20000\,\mathrm{\AA}$,
                                                                vacuum for $20000\,\mathrm{\AA} <   \lambda$) are deduced from experimental energy levels.}
                             $^{(b)}$\,Experimental energy levels taken from \citet{churilovjoshi1995}}.\\
\end{longtable}

\begin{longtable}{rrccrrccrrr}
\caption{\label{tab:sriv:loggf}Calculated HFR oscillator strengths ($\log g_\mathrm{i}f_\mathrm{ik}$) and transition probabilities ($g_\mathrm{k}A_\mathrm{ki}$) 
                               in \ion{Sr}{iv}. CF is the
                               absolute value of the cancellation factor as defined by \citet{cowan1981}. 
                               In Cols\@. 3 and 6, e represents even and o odd.} \\
\hline\hline
\noalign{\smallskip}
                 & \multicolumn{3}{c}{Lower level} && \multicolumn{3}{c}{Upper level} &           &      &    \\
\cline{2-4}\cline{6-8}
\noalign{\smallskip}
Wavelength{$^a$} / \AA & Energy{$^b$} / cm$^{-1}$ & Parity & $J$  && Energy{$^b$} / cm$^{-1}$ & Parity & $J$  & $\log g_\mathrm{i}f_\mathrm{ik}$ & $g_\mathrm{k}A_\mathrm{ki}$ / s$^{-1}$  & CF \\
\noalign{\smallskip}
\hline
\endfirsthead
\caption{continued.}\\
\hline\hline
\noalign{\smallskip}
                 & \multicolumn{3}{c}{Lower level} && \multicolumn{3}{c}{Upper level} &           &      &    \\
\cline{2-4}\cline{6-8}
\noalign{\smallskip}
Wavelength{$^a$} / \AA & Energy{$^b$} / cm$^{-1}$ & Parity & $J$  && Energy{$^b$} / cm$^{-1}$ & Parity & $J$  & $\log g_\mathrm{i}f_\mathrm{ik}$ & $g_\mathrm{k}A_\mathrm{ki}$ / s$^{-1}$  & CF \\
\noalign{\smallskip}
\hline
\noalign{\smallskip}
\endhead
\hline
\noalign{\smallskip}
\endfoot
\noalign{\smallskip}
\hline
\noalign{\smallskip}
\multicolumn{11}{p{15.4cm}}{{Notes.\,
                             $^{(a)}$\,All wavelengths (given in vacuum for                         $\lambda <   2000\,\mathrm{\AA}$, 
                                                                   air for  $2000\,\mathrm{\AA} \le \lambda \le 20000\,\mathrm{\AA}$,
                                                                vacuum for $20000\,\mathrm{\AA} <   \lambda$) are deduced from experimental energy levels.}
                             $^{(b)}$\,Experimental energy levels taken from 
                             \citet{sansonetti2012}.}\\
\end{longtable}

\begin{longtable}{rrccrrccrrr}
\caption{\label{tab:srv:loggf}Calculated HFR oscillator strengths ($\log g_\mathrm{i}f_\mathrm{ik}$) and transition probabilities ($g_\mathrm{k}A_\mathrm{ki}$) 
                               in \ion{Sr}{v}. CF is the
                               absolute value of the cancellation factor as defined by \citet{cowan1981}. 
                               In Cols\@. 3 and 6, e represents even and o odd.} \\
\hline\hline
\noalign{\smallskip}
                 & \multicolumn{3}{c}{Lower level} && \multicolumn{3}{c}{Upper level} &           &      &    \\
\cline{2-4}\cline{6-8}
\noalign{\smallskip}
Wavelength{$^a$} / \AA & Energy{$^b$} / cm$^{-1}$ & Parity & $J$  && Energy{$^b$} / cm$^{-1}$ & Parity & $J$  & $\log g_\mathrm{i}f_\mathrm{ik}$ & $g_\mathrm{k}A_\mathrm{ki}$ / s$^{-1}$  & CF \\
\noalign{\smallskip}
\hline
\endfirsthead
\caption{continued.}\\
\hline\hline
\noalign{\smallskip}
                 & \multicolumn{3}{c}{Lower level} && \multicolumn{3}{c}{Upper level} &           &      &    \\
\cline{2-4}\cline{6-8}
\noalign{\smallskip}
Wavelength{$^a$} / \AA & Energy{$^b$} / cm$^{-1}$ & Parity & $J$  && Energy{$^b$} / cm$^{-1}$ & Parity & $J$  & $\log g_\mathrm{i}f_\mathrm{ik}$ & $g_\mathrm{k}A_\mathrm{ki}$ / s$^{-1}$  & CF \\
\noalign{\smallskip}
\hline
\noalign{\smallskip}
\endhead
\hline
\noalign{\smallskip}
\endfoot
\noalign{\smallskip}
\hline
\noalign{\smallskip}
\multicolumn{11}{p{15.4cm}}{{Notes.\,
                             $^{(a)}$\,All wavelengths (given in vacuum for                         $\lambda <   2000\,\mathrm{\AA}$, 
                                                                   air for  $2000\,\mathrm{\AA} \le \lambda \le 20000\,\mathrm{\AA}$,
                                                                vacuum for $20000\,\mathrm{\AA} <   \lambda$) are deduced from experimental energy levels.}
                             $^{(b)}$\,Experimental energy levels taken from 
                             \citet{sansonetti2012}.}\\
\end{longtable}

\begin{longtable}{rrccrrccrrr}
\caption{\label{tab:srvi:loggf}Calculated HFR oscillator strengths ($\log g_\mathrm{i}f_\mathrm{ik}$) and transition probabilities ($g_\mathrm{k}A_\mathrm{ki}$) 
                               in \ion{Sr}{vi}. CF is the  
                               absolute value of the cancellation factor as defined by \citet{cowan1981}. 
                               In Cols\@. 3 and 6, e represents even and
o odd.} \\
\hline\hline
\noalign{\smallskip}
                 & \multicolumn{3}{c}{Lower level} && \multicolumn{3}{c}{Upper level} &           &      &    \\
\cline{2-4}\cline{6-8}
\noalign{\smallskip}
Wavelength{$^a$} / \AA & Energy{$^b$} / cm$^{-1}$ & Parity & $J$  && Energy{$^b$} / cm$^{-1}$ & Parity & $J$  & $\log g_\mathrm{i}f_\mathrm{ik}$ & $g_\mathrm{k}A_\mathrm{ki}$ / s$^{-1}$  & CF \\
\noalign{\smallskip}
\hline
\endfirsthead
\caption{continued.}\\
\hline\hline
\noalign{\smallskip}
                 & \multicolumn{3}{c}{Lower level} && \multicolumn{3}{c}{Upper level} &           &      &    \\
\cline{2-4}\cline{6-8}
\noalign{\smallskip}
Wavelength{$^a$} / \AA & Energy{$^b$} / cm$^{-1}$ & Parity & $J$  && Energy{$^b$} / cm$^{-1}$ & Parity & $J$  & $\log g_\mathrm{i}f_\mathrm{ik}$ & $g_\mathrm{k}A_\mathrm{ki}$ / s$^{-1}$  & CF \\
\noalign{\smallskip}
\hline
\noalign{\smallskip}
\endhead
\hline
\noalign{\smallskip}
\endfoot
\noalign{\smallskip}
\hline
\noalign{\smallskip}
\multicolumn{11}{p{15.4cm}}{{Notes.\,
                             $^{(a)}$\,All wavelengths (given in vacuum for                         $\lambda <   2000\,\mathrm{\AA}$, 
                                                                   air for  $2000\,\mathrm{\AA} \le \lambda \le 20000\,\mathrm{\AA}$,
                                                                vacuum for $20000\,\mathrm{\AA} <   \lambda$) are deduced from experimental energy levels.}
                             $^{(b)}$\,Experimental energy levels taken from 
                             \citet{sansonetti2012}.}\\
\end{longtable}

\begin{longtable}{rrccrrccrrr}
\caption{\label{tab:srvii:loggf}Calculated HFR oscillator strengths ($\log g_\mathrm{i}f_\mathrm{ik}$) and transition probabilities ($g_\mathrm{k}A_\mathrm{ki}$) 
                               in \ion{Sr}{vii}. CF is the
                               absolute value of the cancellation factor as defined by \citet{cowan1981}. 
                               In Cols\@. 3 and 6, e represents even and
o odd.} \\
\hline\hline
\noalign{\smallskip}
                 & \multicolumn{3}{c}{Lower level} && \multicolumn{3}{c}{Upper level} &           &      &    \\
\cline{2-4}\cline{6-8}
\noalign{\smallskip}
Wavelength{$^a$} / \AA & Energy{$^b$} / cm$^{-1}$ & Parity & $J$  && Energy{$^b$} / cm$^{-1}$ & Parity & $J$  & $\log g_\mathrm{i}f_\mathrm{ik}$ & $g_\mathrm{k}A_\mathrm{ki}$ / s$^{-1}$  & CF \\
\noalign{\smallskip}
\hline
\endfirsthead
\caption{continued.}\\
\hline\hline
\noalign{\smallskip}
                 & \multicolumn{3}{c}{Lower level} && \multicolumn{3}{c}{Upper level} &           &      &    \\
\cline{2-4}\cline{6-8}
\noalign{\smallskip}
Wavelength{$^a$} / \AA & Energy{$^b$} / cm$^{-1}$ & Parity & $J$  && Energy{$^b$} / cm$^{-1}$ & Parity & $J$  & $\log g_\mathrm{i}f_\mathrm{ik}$ & $g_\mathrm{k}A_\mathrm{ki}$ / s$^{-1}$  & CF \\
\noalign{\smallskip}
\hline
\noalign{\smallskip}
\endhead
\hline
\noalign{\smallskip}
\endfoot
\noalign{\smallskip}
\hline
\noalign{\smallskip}
\multicolumn{11}{p{15.4cm}}{{Notes.\,
                             $^{(a)}$\,All wavelengths (given in vacuum for                         $\lambda <   2000\,\mathrm{\AA}$, 
                                                                   air for  $2000\,\mathrm{\AA} \le \lambda \le 20000\,\mathrm{\AA}$,
                                                                vacuum for $20000\,\mathrm{\AA} <   \lambda$) are deduced from experimental energy levels.}
                             $^{(b)}$\,Experimental energy levels taken from \citet{sansonetti2012}.}\\
\end{longtable}

\begin{longtable}{rrccrrccrrr}
\caption{\label{tab:tevi:loggf}Calculated HFR oscillator strengths ($\log g_\mathrm{i}f_\mathrm{ik}$) and transition probabilities ($g_\mathrm{k}A_\mathrm{ki}$) 
                               in \ion{Te}{vi}. CF is the
                               absolute value of the cancellation factor as defined by \citet{cowan1981}. 
                               In Cols\@. 3 and 6, e represents even and
o odd.} \\
\hline\hline
\noalign{\smallskip}
                 & \multicolumn{3}{c}{Lower level} && \multicolumn{3}{c}{Upper level} &           &      &    \\
\cline{2-4}\cline{6-8}
\noalign{\smallskip}
Wavelength{$^a$} / \AA & Energy{$^b$} / cm$^{-1}$ & Parity & $J$  && Energy{$^b$} / cm$^{-1}$ & Parity & $J$  & $\log g_\mathrm{i}f_\mathrm{ik}$ & $g_\mathrm{k}A_\mathrm{ki}$ / s$^{-1}$  & CF \\
\noalign{\smallskip}
\hline
\endfirsthead
\caption{continued.}\\
\hline\hline
\noalign{\smallskip}
                 & \multicolumn{3}{c}{Lower level} && \multicolumn{3}{c}{Upper level} &           &      &    \\
\cline{2-4}\cline{6-8}
\noalign{\smallskip}
Wavelength{$^a$} / \AA & Energy{$^b$} / cm$^{-1}$ & Parity & $J$  && Energy{$^b$} / cm$^{-1}$ & Parity & $J$  & $\log g_\mathrm{i}f_\mathrm{ik}$ & $g_\mathrm{k}A_\mathrm{ki}$ / s$^{-1}$  & CF \\
\noalign{\smallskip}
\hline
\noalign{\smallskip}
\endhead
\hline
\noalign{\smallskip}
\endfoot
\noalign{\smallskip}
\hline
\noalign{\smallskip}
\multicolumn{11}{p{15.4cm}}{{Notes.\,
                             $^{(a)}$\,All wavelengths (given in vacuum for                         $\lambda <   2000\,\mathrm{\AA}$, 
                                                                   air for  $2000\,\mathrm{\AA} \le \lambda \le 20000\,\mathrm{\AA}$,
                                                                vacuum for $20000\,\mathrm{\AA} <   \lambda$) are deduced from experimental energy levels.}
                             $^{(b)}$\,Experimental energy levels taken from 
                             \citet{crookerjoshi1964}, 
                             \citet{dunneosullivan1992}, and 
                             \citet{ryabtsevetal2007}.}\\
\end{longtable}

\begin{longtable}{rrccrrccrrr}
\caption{\label{tab:ivi:loggf}Calculated HFR oscillator strengths ($\log g_\mathrm{i}f_\mathrm{ik}$) and transition probabilities ($g_\mathrm{k}A_\mathrm{ki}$) 
                               in \ion{I}{vi}. CF is the
                               absolute value of the cancellation factor as defined by \citet{cowan1981}. 
                               In Cols\@. 3 and 6, e represents even and
o odd.} \\
\hline\hline
\noalign{\smallskip}
                 & \multicolumn{3}{c}{Lower level} && \multicolumn{3}{c}{Upper level} &           &      &    \\
\cline{2-4}\cline{6-8}
\noalign{\smallskip}
Wavelength{$^a$} / \AA & Energy{$^b$} / cm$^{-1}$ & Parity & $J$  && Energy{$^b$} / cm$^{-1}$ & Parity & $J$  & $\log g_\mathrm{i}f_\mathrm{ik}$ & $g_\mathrm{k}A_\mathrm{ki}$ / s$^{-1}$  & CF \\
\noalign{\smallskip}
\hline
\endfirsthead
\caption{continued.}\\
\hline\hline
\noalign{\smallskip}
                 & \multicolumn{3}{c}{Lower level} && \multicolumn{3}{c}{Upper level} &           &      &    \\
\cline{2-4}\cline{6-8}
\noalign{\smallskip}
Wavelength{$^a$} / \AA & Energy{$^b$} / cm$^{-1}$ & Parity & $J$  && Energy{$^b$} / cm$^{-1}$ & Parity & $J$  & $\log g_\mathrm{i}f_\mathrm{ik}$ & $g_\mathrm{k}A_\mathrm{ki}$ / s$^{-1}$  & CF \\
\noalign{\smallskip}
\hline
\noalign{\smallskip}
\endhead
\hline
\noalign{\smallskip}
\endfoot
\noalign{\smallskip}
\hline
\noalign{\smallskip}
\multicolumn{11}{p{15.4cm}}{{Notes.\,
                             $^{(a)}$\,All wavelengths (given in vacuum for                         $\lambda <   2000\,\mathrm{\AA}$, 
                                                                   air for  $2000\,\mathrm{\AA} \le \lambda \le 20000\,\mathrm{\AA}$,
                                                                vacuum for $20000\,\mathrm{\AA} <   \lambda$) are deduced from experimental energy levels.}
                             $^{(b)}$\,Experimental energy levels taken from 
                             \citet{tauheedetal1997}.}\\
\end{longtable}

\clearpage

\section{Additional figures}
\label{app:addfigs}

\begin{figure*}[h!]
   \resizebox{\hsize}{!}{\includegraphics{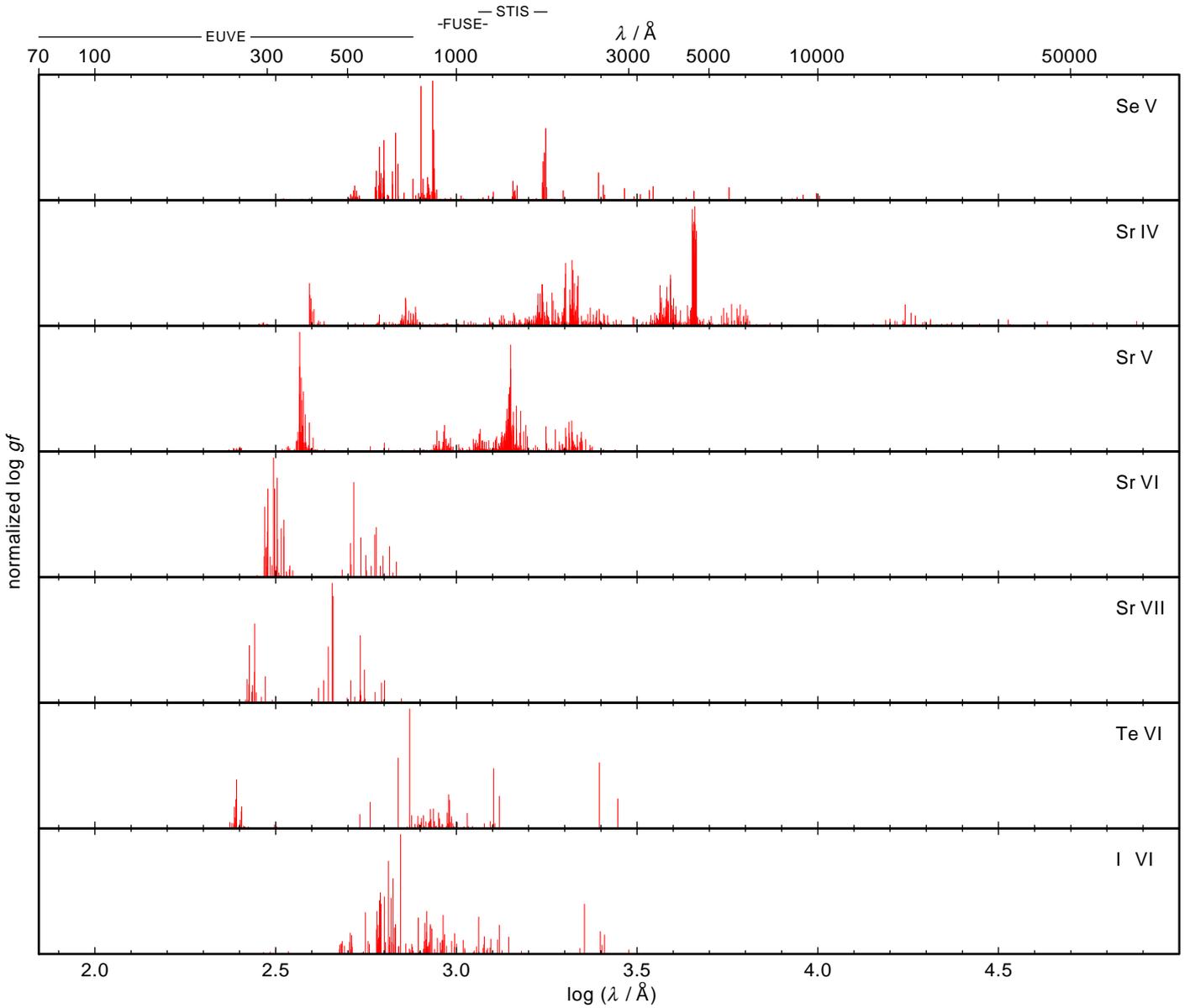}}
    \caption{Newly calculated $\log g_\mathrm{i}f_\mathrm{ik}$ values of \ion{Se}{v}, \ion{Sr}{iv - vii},  \ion{Te}{vi}, and  \ion{I}{vi}
             (from top to bottom). The $\log g_\mathrm{i}f_\mathrm{ik}$ values are normalized to the strongest line, matching 95\,\% of
             the panels' heights. The wavelength ranges of EUVE and of our FUSE and STIS spectra are indicated at
             the top.
            }
   \label{fig:normloggf}
\end{figure*}

\twocolumn
\end{appendix}

\end{document}